\documentclass[aps,prd,preprint,draft,showpacs,eqsecnum,nofootinbib]{revtex4}
\usepackage{graphicx,epsf,amssymb}
\newcommand{\be}{\begin{equation}}
\newcommand{\ee}{\end{equation}}
\newcommand{\ba}{\begin{eqnarray}}
\newcommand{\ea}{\end{eqnarray}}

\newcommand{\bea}{\begin{eqnarray}}
\newcommand{\eea}{\end{eqnarray}}

\newif\ifdraft
\drafttrue
\newif\ifpreprint
\preprinttrue

\def\sandp#1.#2.#3{%
\left\langle\smash{#1}{\vphantom1}^{-}\right|{#2}%
\left|\smash{#3}{\vphantom1}^{+}\right\rangle}
\def\sandpp#1.#2.#3{%
\left\langle\smash{#1}{\vphantom1}^{+}\right|{#2}%
\left|\smash{#3}{\vphantom1}^{+}\right\rangle}
\def\sandmm#1.#2.#3{%
\left\langle\smash{#1}{\vphantom1}^{-}\right|{#2}%
\left|\smash{#3}{\vphantom1}^{-}\right\rangle}
\def\spab#1.#2.#3{\sandmm#1.#2.#3}
\def\spba#1.#2.#3{\sandpp#1.#2.#3}
\def\spaa#1.#2.#3.#4{\sandmp#1.{#2#3}.#4}
\def\spbb#1.#2.#3.#4{\sandpm#1.{#2#3}.#4}
\def\spa#1.#2{\langle#1\,#2\rangle}
\def\spb#1.#2{[#1\,#2]}
\def\spash#1.#2{\vphantom{\hat K}\spa{\smash{#1}}.{\smash{#2}}}
\def\spbsh#1.#2{\vphantom{\hat K}\spb{\smash{#1}}.{\smash{#2}}}
\def\lor#1.#2{\left(#1\,#2\right)}
\def\sand#1.#2.#3{%
\left\langle\smash{#1}{\vphantom1}^{-}\right|{#2}%
\left|\smash{#3}{\vphantom1}^{-}\right\rangle}
\def\sandpp#1.#2.#3{%
\left\langle\smash{#1}{\vphantom1}^{+}\right|{#2}%
\left|\smash{#3}{\vphantom1}^{+}\right\rangle}
\def\sandpm#1.#2.#3{%
\left\langle\smash{#1}{\vphantom1}^{+}\right|{#2}%
\left|\smash{#3}{\vphantom1}^{-}\right\rangle}
\def\sandmp#1.#2.#3{%
\left\langle\smash{#1}{\vphantom1}^{-}\right|{#2}%
\left|\smash{#3}{\vphantom1}^{+}\right\rangle}

\newbox\SlashedBox
\def\slashed#1{\setbox\SlashedBox=\hbox{#1}
\hbox to 0pt{\hbox to 1\wd\SlashedBox{\hfil/\hfil}\hss}#1}
\def\hboxtosizeof#1#2{\setbox\SlashedBox=\hbox{#1}
\hbox to 1\wd\SlashedBox{#2}}

\newbox\charbox
\newbox\slabox
\def\s#1{{      
        \setbox\charbox=\hbox{$#1$}
        \setbox\slabox=\hbox{$/$}
        \dimen\charbox=\ht\slabox
        \advance\dimen\charbox by -\dp\slabox
        \advance\dimen\charbox by -\ht\charbox
        \advance\dimen\charbox by \dp\charbox
        \divide\dimen\charbox by 2
        \raise-\dimen\charbox\hbox to \wd\charbox{\hss/\hss}
        \llap{$#1$}
}}

\def\eqn#1{eq.~(\ref{#1})}

\def\qb{{\bar q}}

\def\sign{{\mathop{\rm sign}\nolimits}}

\def\tree{{\rm tree}}

\def\sandp#1.#2.#3{%
\left\langle\smash{#1}{\vphantom1}^{+}\right|{#2}%
\left|\smash{#3}{\vphantom1}^{+}\right\rangle}

\def\Den#1#2 {\prod\limits_{k=#1}^{#2} \spa{k}.{(k+1)}}

\def\Res{\mathop{\rm Res}}
\def\tlambda{{\tilde\lambda}}

\def\Res{\mathop{\rm Res}}

\newcommand{\Bmp}[1]{\langle #1\rangle}

\newcommand{\At}{A^{\tree}}

\newbox\ourfigbox
\def\SizedFigureWithCaption#1#2#3{%
\setbox\ourfigbox
  \hbox{\hss\epsfxsize #1 \epsfbox{#2}\hss}
\hbox to \wd\ourfigbox{\vbox{\noindent\copy\ourfigbox\break
\vskip -6mm      \hbox to \wd\ourfigbox{\hss#3\hss}}}
}

\def\llongrightarrow{%
\relbar\mskip-0.5mu\joinrel\mskip-0.5mu\relbar
     \mskip-0.5mu\joinrel\longrightarrow}
\def\inlimit^#1{\buildrel#1\over\llongrightarrow}
\def\dash{\hbox{-\kern-.02em}}

\newcommand{\Inf}{$\textrm{Inf}$}

\newcommand{\Kf}[1]{K^{\flat}_{#1}}
\newcommand{\Kfp}[1]{K^{\flat,+}_{#1}}
\newcommand{\Kfm}[1]{K^{\flat,-}_{#1}}

\newcommand{\Kfpm}[1]{K^{\flat,\pm}_{#1}}

\newcommand{\qbd}{\overline{q}}
\newcommand{\ebd}{\overline{e}}

\begin{document}
\hfuzz 10 pt


\ifpreprint
\noindent
SLAC--PUB--12455
\hfill UCLA/07/TEP/12
\fi

\title{Direct extraction of one-loop integral coefficients
\footnote{Research supported in part by the US Department of
 Energy under contracts DE--FG03--91ER40662 and DE--AC02--76SF00515.}
}

\author{Darren Forde}
\affiliation{Stanford Linear Accelerator Center \\
             Stanford University\\
             Stanford, CA 94309, USA,\\
and\\
 Department of Physics and Astronomy, UCLA\\
\hbox{Los Angeles, CA 90095--1547, USA.}
}

\date{$12^{\textrm{th}}$ April 2007}

\begin{abstract}
We present a general procedure for obtaining the coefficients of the
scalar bubble and triangle integral functions of one-loop
amplitudes. Coefficients are extracted by considering two-particle and
triple unitarity cuts of the corresponding bubble and triangle
integral functions. After choosing a specific parameterisation of the
cut loop momentum we can uniquely identify the coefficients of the
desired integral functions simply by examining the behaviour of the
cut integrand as the unconstrained parameters of the cut loop momentum
approach infinity. In this way we can produce compact forms for scalar
integral coefficients. Applications of this method are presented for
both QCD and electroweak processes, including an alternative form for
the recently computed three-mass triangle coefficient in the
six-photon amplitude $A_6(1^-,2^+,3^-,4^+,5^-,6^+)$. The direct nature
of this extraction procedure allows for a very straightforward
automation of the procedure.
\end{abstract}

\pacs{11.15.Bt, 11.25.Db, 12.15.Lk, 12.38.Bx \hspace{1cm}}

\maketitle


\renewcommand{\thefootnote}{\arabic{footnote}}
\setcounter{footnote}{0}

\section{Introduction}
\label{IntroSection}

Maximising the discovery potential of future colliders such as CERN's
Large Hadron Collider (LHC) will rely upon a detailed understanding of
Standard Model processes. Distinguishing signals of new physics from
background processes requires precise theoretical calculations. These
background processes need to be known to at least a next-to-leading
order (NLO) level. This in turn entails the need for computation of
one-loop amplitudes. Whilst much progress has been made in calculating
such processes, the feasibility of producing these needed higher
multiplicity amplitudes, such as one-loop processes with one or more
vector bosons (W's, Z's and photons) along with multiple jets, strains
standard Feynman diagram techniques.

Direct calculations using Feynman diagrams are generally inefficient;
the large number of terms and diagrams involved has by necessity
demanded (semi)numerical approaches be taken when dealing with higher
multiplicity amplitudes. Much progress has been made in this way,
numerical evaluations of processes with up to six partons have been
performed~\cite{Denner,GieleGloverNumerical,EGZ,EGZ06,OtherNumerical}. On
assembling complete amplitudes from Feynman diagrams it is commonly
found that large cancellations take place between the various
terms. The remaining result is then far more compact than would
naively be expected from the complexity of the original Feynman
diagrams. The greater simplicity of these final forms has spurred the
development of alternative more direct and efficient techniques for
calculating these processes.

The elegant and efficient approach of recursion relations has long
been a staple part of the tree level calculational
approach~\cite{BGRecurrence,DAKRecurrence}. Recent progress, inspired
by developments in twistor string
theory~\cite{WittenTopologicalString,CSW}, builds upon the idea of
recursion relations, but centred around the use of gauge-independent
or on-shell intermediate quantities and hence negating a potential
source of large cancellations between terms. Britto, Cachazo and
Feng~\cite{Britto:2004ap} initially wrote down a set of tree level
recursion relations utilising \textit{on-shell} amplitudes with
\textit{complex} values of external momenta. Then, along with
Witten~\cite{Britto:2005fq}, they proved these on-shell recursion
relations using just a knowledge of the factorisation properties of
the amplitudes and Cauchy's theorem. The generality of the proof has
led to their application in many diverse areas beyond that of massless
gluons and fermions in gauge
theory~\cite{Britto:2004ap,TreeRecurResults}. There have been
extensions to theories with massive scalars and
fermions~\cite{GloverMassive,Massive,Schwinn:2007ee} as well as
amplitudes in gravity~\cite{Gravity}.

Similarly ``on-shell'' approaches can also be constructed at loop
level. The unitarity of the perturbative $S$-matrix can be used to
produce compact analytical results by ``gluing'' together on-shell
tree amplitudes to form the desired loop amplitude. This unitarity
approach has been developed into a practical technique for the
construction of loop
amplitudes~\cite{Neq4Oneloop,Neq4Oneloop2,Bern:1997sc}, initially, for
computational reasons, for the construction of amplitudes where the
loop momentum was kept in $D=4$ dimensions. This limited its
applicability to computations of the ``cut-constructible'' parts of an
amplitude only, i.e. (poly)logarithmic containing terms and any
associated $\pi^2$ constants. Amplitudes consisting of only such
terms, such as supersymmetric amplitudes, can therefore be completely
constructed in this way. QCD amplitudes contain in addition rational
pieces which cannot be derived using such cuts. The ``missing''
rational parts are constructible directly from the unitarity approach
only by taking the cut loop momentum to be in $D=4-2\epsilon$
dimensions~\cite{vanNeerven:1985xr}. The greater difficulty of such
calculations has, with only a few
exceptions~\cite{BernMorgan,BMSTUnitarity}, restricted the application
of this approach, although recent
developments~\cite{Anastasiou:2006jv,Anastasiou:2006gt,Britto:2006fc}
have provided new promise for this direction.

The generality of the foundation of on-shell recursion relation
techniques does not limit their applicability to tree level processes
only. The ``missing'' rational pieces at one-loop, in QCD and other
similar theories, can be constructed in an analogous way to (rational)
tree level amplitudes~\cite{Bern:2005hs,Bern:2005cq}. The ``unitarity
on-shell bootstrap'' technique combines unitarity with on-shell
recursion, and provides, in an efficient manner, the complete one-loop
amplitude.  This approach has been used to produce various new
analytic results for amplitudes containing both fixed numbers as well
as arbitrary numbers of external
legs~\cite{Berger:2006vq,Forde:2005hh,Berger:2006ci}. Other newly
developed alternative methods have also proved fruitful for
calculating rational
terms~\cite{Xiao:2006vr,Su:2006vs,Xiao:2006vt,Binoth:2006hk}. In
combination with the required cut-containing
terms~\cite{Britto:2005ha,Britto:2006sj,BBSTQCD} these new results for
the rational loop contributions combine to give the complete analytic
form for the one-loop QCD six-gluon amplitude.

The development of efficient techniques for calculating, what were
previously difficult to derive rational terms, has emphasised the need
to optimise the derivation of the cut-constructible pieces of the
amplitude. One-loop amplitudes can be decomposed entirely in terms of
a basis of scalar bubble, scalar triangle and scalar box integral
functions. Deriving cut-constructible terms therefore reduces to the
problem of finding the coefficients of these basis integrals. For the
coefficients of scalar box integrals it was shown
in~\cite{Britto:2004nc} that a combination of \textit{generalised}
unitarity~\cite{Bern:1997sc,TwoLoopSplit,NeqFourSevenPoint,Eden},
quadruple cuts in this case, along with the use of \textit{complex}
momenta could be used, within a purely algebraic approach, to extract
the desired coefficient from the cut integrand of the associated box
topology.

Extracting triangle and bubble coefficients presents more of a
problem. Unlike for the case of box coefficients, cutting all the
propagators associated with the desired integral topology does not
uniquely isolate a single integral coefficient. Inside a particular
two-particle or triple cut lie multiple scalar integral coefficients
corresponding to integrals with topologies sharing not only the same
cuts but also additional propagators. These coefficients must
therefore be disentangled in some way. There are multiple directions
within the literature which have been taken to effect this
separation. The pioneering work by Bern, Dixon, Dunbar and Kosower
related unitarity cuts to Feynman diagrams and thence to the scalar
integral basis, this then allowed for the derivation of many important
results~\cite{Neq4Oneloop,Neq4Oneloop2,Bern:1997sc}. More recently the
technique of Britto
\textit{et. al.}~\cite{Britto:2005ha,Britto:2006sj,
Anastasiou:2006jv,Anastasiou:2006gt,Britto:2006fc} has for
two-particle cuts and the its extension to triple cuts by
Mastrolia~\cite{Mastrolia:2006ki}, highlighted the benefits of working
in a spinor formalism, where the cut integrals can be integrated
directly. Important results obtained in this way include the most
difficult of the cut-constructable pieces for the one-loop amplitude
for six gluons with the helicity configurations $A_6(+-+-+-)$ and
$A_6(-+--++)$. The cut-constructible parts of
Maximum-Helicity-Violating (MHV) one-loop amplitudes were found by
joining MHV amplitudes together in a similar manner to at tree
level~\cite{BST}. This method has been applied by Bedford, Brandhuber,
Spence and Travaglini to produce new QCD results~\cite{BBSTQCD}. In
the approach of Ossola, Papadopoulos and
Pittau~\cite{Ossola:2006us,Ossola:2007bb} it is possible to avoid the
need to perform any integration or use any integral reduction
techniques. Coefficients are instead extracted by solving sets of
equations. The solutions of these equations include the desired
coefficients, along with additional ``spurious'' terms corresponding
to coefficients of terms which vanish after integrating over the loop
momenta.

The many-fold different processes and their differing parton contents
that will be needed at current and future collider experiments
suggests that some form of automation, even of the more efficient
``on-shell'' techniques, will be required. From an efficiency
standpoint, therefore, we would ideally wish to minimise the degree of
calculation required for each step of any such process. Here we
propose a new method for the extraction of scalar integral
coefficients which aims to meet this goal. The technique follows in
the spirit of the simplicity of the derivation of scalar box
coefficients given in ref.~\cite{Britto:2004nc}. Desired coefficients can
be \textit{constructed} directly using two-particle or triple cuts.
The complete one-loop amplitude can then be obtained by summing over
all such cuts and adding any box terms and rational
pieces. Alternatively our technique can be used to \textit{extract}
the bubble and triangle coefficients from a one-loop amplitude,
generated for example from a Feynman diagram. Hence the technique is
acting as an efficient way to perform the integration.

We use unitarity cuts to freeze some of the degrees of freedom of the
integral loop momentum, whilst leaving others unconstrained. This then
isolates a specific single bubble or triangle integral topology and
hence its coefficient. Within each cut there remain additional
coefficients. In the triangle case those of scalar box integrals.  In
the bubble case both scalar box and scalar triangle integrals
contribute. Disentangling our desired coefficient from these extra
contributions is a straightforward two step procedure. First one
rewrites the loop momentum inside the cut integrand in terms of its
unconstrained parameters. In the triangle case there is a single
parameter, and in the bubble case there are a pair of
parameters. Examining the behaviour of the integrand as these
unconstrained parameters approach infinity then allows for a
straightforward separation of the desired coefficient from any extra
contributions. The coefficient of each basis integral function can
therefore be extracted individually in an efficient manner with no
further computation.

This paper is organised as follows. In section~\ref{NotationSection}
we outline the notation used throughout this paper. In
section~\ref{section:unitarity_cutting_techniques} we proceed to
present the basic structure of a one-loop amplitude in terms of a
basis of scalar integral functions. We describe in
section~\ref{eq:triple_cuts_and_scalar_triangles} our procedure for
extracting the coefficients of scalar triangle coefficients through
the use of a particular loop-momentum parameterisation for the triple
cuts along with the properties of the cut as the single free integral
parameter tends to infinity.
Section~\ref{section:double_cuts_and_bubble_coeffs} extends this
formalism to include the extraction of scalar bubble coefficients.
The two-particle cut used in this case contains an additional free
parameter and requires an additional step in our procedure. Finally in
section~\ref{section:examples} we conclude by providing some
applications which act as checks of our method. Initially we examine
the extraction of various basis integral coefficients from some common
one-loop integral functions. We then turn our attention to the
construction of the coefficients of some more phenomenologically
interesting processes. These include the three-mass triangle
coefficient for the six photon amplitude $A_6(-+-+-+)$, as well as a
representative three-mass triangle coefficient of the process
$e^+e^-\rightarrow q^+\overline{q}^-g^-g^+$. Finally we construct the
complete cut-containing part of the amplitude $A^{\rm 1-loop
}_6(1^-,2^-,3^+,4^+,5^+)$ and discuss further comparisons against
coefficients of more complicated gluon amplitudes contained in the
literature.


\section{Notation}
\label{NotationSection}

In this section we summarise the notation used in the remainder of the
paper. We will use the spinor helicity
formalism~\cite{SpinorHelicity,TreeReview}, in which the amplitudes
are expressed in terms of spinor inner-products,
\begin{equation}
\spa{j}.{l} = \langle j^- | l^+ \rangle = \bar{u}_-(k_j) u_+(k_l)\,,
\hskip 2 cm
\spb{j}.{l} = \langle j^+ | l^- \rangle = \bar{u}_+(k_j) u_-(k_l)\, ,
\label{spinorproddef}
\end{equation}
where $u_\pm(k)$ is a massless Weyl spinor with momentum $k$ and
positive or negative chirality. The notation used here follows the QCD
literature, with $\spb{i}.{j} = \sign(k_i^0 k_j^0)\spa{j}.{i}^*$ for
real momenta so that,
\begin{equation}
\spa{i}.{j} \spb{j}.{i} = 2 k_i \cdot k_j = s_{ij}\,.
\end{equation}
Our convention is that all legs are outgoing.
We also define,
\begin{equation}
 \lambda_i \equiv u_+(k_i), \qquad \tlambda_i \equiv u_-(k_i) \,.
\label{lambdadef}
\end{equation}

We denote the sums of cyclicly-consecutive external momenta by
\begin{equation}
K^\mu_{i\ldots j} \equiv
   k_i^\mu + k_{i+1}^\mu + \cdots + k_{j-1}^\mu + k_j^\mu \,,
\label{KDef}
\end{equation}
where all indices are mod $n$ for an $n$-gluon amplitude.
The invariant mass of this vector is
\be
s_{i\ldots j} \equiv K_{i\ldots j}^2\,.
\ee
Special cases include the two- and three-particle invariant masses,
which are denoted by
\begin{equation}
s_{ij} \equiv K_{ij}^2
\equiv (k_i+k_j)^2 = 2k_i\cdot k_j,
\qquad \quad
s_{ijk} \equiv (k_i+k_j+k_k)^2 \,.
\label{TwoThreeMassInvariants}
\end{equation}
We also define spinor strings,
\begin{eqnarray}
\spab{i}.{(\s a\pm \s b)}.{j} &=& \spa{i}.{a} \spb{a}.{j} \pm \spa{i}.{b} \spb{b}.{j} \,,
          \nonumber   \\
\spbb{i}.{(\s a+\s b)}.{(\s c+\s d)}.{j} &=&
     \spb{i}.{a} \spab{a}.{(\s c+\s d)}.{j} +
     \spb{i}.{b} \spab{b}.{(\s c+\s d)}.{j} \,.
\end{eqnarray}


\section{Unitarity cutting techniques and the one-loop integral basis}
\label{section:unitarity_cutting_techniques}

Our starting point will be the general dimensionally-regularised
decomposition of a one-loop amplitude into a basis of scalar integral
functions~\cite{Neq4Oneloop2,Bern:1993kr}
\begin{eqnarray}
A^{\rm 1-loop}_n\!=\!\mathcal{R}_n\!+\!r_{\Gamma}\frac{(\mu^2)^{\epsilon}}{(4\pi)^{2-\epsilon}}\!\left(\!\sum_i
\!b_i B_0(K^2_i)\!+\!\sum_{ij}\!c_{ij}
C_0(K^2_i,K^2_j)\!+\!\sum_{ijk}\!d_{ijk}D_0(K^2_i,K^2_j,K^2_k)\!\right).\label{eq:basis_def}
\end{eqnarray}
The scalar bubble, triangle and box integral functions are denoted by
$B_0$, $C_0$ and $D_0$ respectively, and along with $r_{\Gamma}$ their
explicit forms can be found in
Appendix~\ref{appendix:scalar_int_fns}. The $b_i$, $c_{ij}$ and
$d_{{ijk}}$ are their corresponding rational coefficients. Any
$\epsilon$ dependence within these coefficients has been removed and
placed into the rational, $\mathcal{R}_n$, term.  The problem of
deriving the one-loop amplitude is therefore reduced to that of
finding the coefficients of these scalar integral functions and any
rational terms when working in $D=4$ dimensions.

We are going to consider obtaining these coefficients via the
application of various cuts within the framework of generalised
unitarity~\cite{Bern:1997sc,TwoLoopSplit,NeqFourSevenPoint,Eden}. In
general our cut momenta will be complex, so for our purposes we define
a ``cut'' as the replacement
\begin{eqnarray}
\frac{i}{(l+K_i)^2}\rightarrow (2\pi)\delta((l+K_i)^2).
\end{eqnarray}
By systematically \textit{constructing} all possible unitarity cuts we
can reproduce every integral coefficient of a particular
amplitude. Alternatively, application of the same procedure of
``cutting'' legs can be used to \textit{extract} from a one-loop
integral the corresponding coefficients of the standard basis
integrals making up that particular integral, in a sense acting as a
form of specialised integral reduction. This approach follows in a
similar vein to that adopted by Ossola, Papadopoulos and
Pittau~\cite{Ossola:2006us}.

The most straightforward implementation of the technique we present
here is when the cut loop momentum is massless and kept in $D=4$
dimensions. Eq.~\ref{eq:basis_def} therefore contains, within the term
$\mathcal{R}_n$, any rational terms missed by performing cuts in only
$D=4$. Approaches for deriving such terms independently of unitarity
cuts exist and so we do not concern ourselves with these
here~\cite{Anastasiou:2006jv,Anastasiou:2006gt,Bern:2005hs,Bern:2005cq,Berger:2006vq,Berger:2006ci,Xiao:2006vr,Su:2006vs,Xiao:2006vt,Binoth:2006hk,Ossola:2006us,Ossola:2007bb}.

\begin{figure}[ht]
\centerline{\epsfxsize 2.7 truein\epsfbox{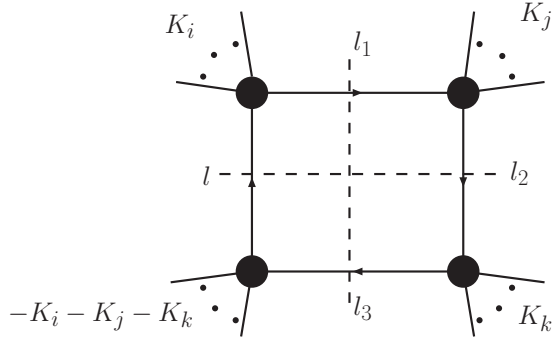}}
\caption{ A generic quadruple cut used to isolate the scalar box
integral $D_0(K^2_i,K^2_j,K^2_k)$.  }
\label{figure:box_coeff}
\end{figure}

As was demonstrated in~\cite{Britto:2004nc}, the application of a
quadruple cut, as shown in figure~\ref{figure:box_coeff},
to $A^{\rm 1-loop}_n$ uniquely identifies a particular box integral
topology $D_0(K^2_i,K^2_j,K^2_k)$ and hence its coefficient. This
coefficient is then given by
\begin{eqnarray}
d_{ijk}=\frac{1}{2}\sum_{a=1}^2A_1(l_{ijk;a})A_2(l_{ijk;a})A_3(l_{ijk;a})A_4(l_{ijk;a}),\label{eq:box_coeffs}
\end{eqnarray}
where $l_{ijk;a}$ is the $a^{\rm th}$ solution of the cut loop
momentum $l$ that isolates the scalar box function
$D_0(K^2_i,K^2_j,K^2_k)$, there are $2$ such solutions.
Eq.~\ref{eq:box_coeffs} applies as well to the cases when one or more
of the four legs of the box is massless. This is a result of the
existence, for complex momenta, of a well-defined three-point tree
amplitude corresponding to any corner of a box containing a massless
leg.

Applying a triple cut to the amplitude $A^{\rm 1-loop}_n$ does not
isolate a single basis integral. Instead we have a triangle integral
plus a sum of box integrals obtained by ``opening'' a fourth
propagator. This can be represented schematically via
\begin{eqnarray}
r_{\Gamma}\frac{(\mu^2)^{\epsilon}}{(4\pi)^{2-\epsilon}}\left(c_{ij}
C_0(K^2_i,K^2_j)+\sum_{k}d_{ijk}D_0(K^2_i,K^2_j,K^2_k)+\ldots\right),
\end{eqnarray}
where the additional terms correspond to ``opening'' the $K_i$ leg or
the $K_j$ leg instead of the $-(K_i+K_j)$ leg.  Similarly in the case
of a two-particle cut we again cannot isolate a single basis integral
by itself. Instead we get additional triangle and box integrals
corresponding to ``opening'' third and forth
propagators. Schematically this is given by
\begin{eqnarray}
r_{\Gamma}\frac{(\mu^2)^{\epsilon}}{(4\pi)^{2-\epsilon}}\left( b_i B_0(K^2_i)+\sum_{j}c_{ij} C_0(K^2_i,K^2_j)+\sum_{jk}d_{ijk}D_0(K^2_i,K^2_j,K^2_k)+\ldots\right),
\end{eqnarray}
where again the additional terms are boxes with the $K_i$ leg or the
$K_j$ legs ``opened''.  Whilst not isolating a single integral each of the
above cuts does single out either \textit{one} scalar triangle, in the
triple cut case, or \textit{one} scalar bubble, in the two-particle
cut case. Disentangling these single bubble or triangle integral
functions from the contributions of the remaining basis integrals
will allow us to directly read off the corresponding
coefficient. Applying all possible two-particle, triple and quadruple
cuts then enables us to derive the coefficients of every basis
integral function.


\section{Triple cuts and scalar triangle coefficients}
\label{eq:triple_cuts_and_scalar_triangles}

A triple cut contains not only contributions for the corresponding
scalar triangle integral, but also contributions from scalar box
integrals which share the same three cuts as the triangle. Of the four
propagators of a scalar box integral, three will be given by the three
cut legs of the triple cut loop integral. The forth propagator will be
contained inside the cut integrand in a denominator factor of the form
$(l-P)^2$, which corresponds to a propagator pole. Ideally we want
to separate terms containing such poles from the remainder of the cut
integrand. The remaining term will be the scalar triangle integral
multiplied by its coefficient for that particular cut.

The three delta functions of a triple cut constrain the cut loop
momentum such that only a single free parameter of the integral
remains, which we label $t$. We can express the loop momentum in terms
of this parameter using the orthogonal null four-vectors, $a^{\mu}_i$,
with $i=1,2,3$, specific forms for these basis vectors are presented
in section~\ref{section:the_triple_cut_mom_param}. The loop momentum
is then given by
\begin{eqnarray}
l^{\mu}=a_0^{\mu}t+\frac{1}{t}a_1^{\mu}+a_2^{\mu}.
\end{eqnarray}
Denominator factors of the cut integrand depending upon the cut loop
momentum, can be written as propagators of the general form,
$(l-P)^2$. When these propagators go on-shell they will correspond to
poles in $t$. These poles will be solutions of the following equation
\begin{eqnarray}
(l-P)^2=0\;\;\;\Rightarrow \;\;\;2(a_0\cdot P)t+2(a_1\cdot
  P)\frac{1}{t}+2(a_2\cdot P)-P^2=0.\label{eq:quadratic_sol}
\end{eqnarray}
If we consider $t$ to be a complex parameter then we can use a partial
fraction decomposition in terms of $t$ to rewrite an arbitrary
triple-cut integral.  For the extraction of integral coefficients we
need only work with integrals in $D=4$ dimensions. We also drop an
overall denominator factor of $1/(2\pi)^4$ which multiplies all
integrals. The partial fraction decomposition is therefore given, in
the case when we have applied a triple cut on the legs $l^2$,
$(l-K_1)^2$ and $(l-K_2)^2$, by
\begin{eqnarray}
&&\hspace*{-0.4cm}\!(2\pi)^3\!\!\int d^4l
\prod_{i=0}^2\delta(l_i^2)A_1A_2A_3 \nonumber\\
&&\hphantom{\!-i\!\!\int} =\!(2\pi)^3\!\!\int d^4l
\prod_{i=0}^2\delta(l_i^2)\!\left(\left[\textrm{Inf}_{t}
A_1A_2A_3\right](t)+\!\!\!\!\!\sum_{\textrm{poles } \{j\}}
\!\!\!\frac{\Res_{t=t_j} A_1A_2A_3}{t-t_j}\right),
\label{eq:on_shell_split}
\end{eqnarray}
where $l_i=l-K_i$ and $l_0=l$. This is a sum of all possible poles of
$t$, labelled here as the set $\{j\}$, contained in the cut integrand
denoted by $A_1A_2A_3$. Pieces of the integrand without a pole are
contained in the \Inf{} term, originally given
in~\cite{Berger:2006ci}, and defined such that
\begin{eqnarray}
\lim_{t\rightarrow \infty}\left(\left[\textrm{Inf}_t A_1A_2A_3\right](t)-A_1(t)A_2(t)A_3(t)\right)=0.
\end{eqnarray}
In general $[\textrm{Inf}_z A_1A_2A_3](t)$ will be some polynomial in
$t$,
\begin{eqnarray}
\left[{\rm Inf}_{t}A_1A_2A_3\right](t)=\sum_{i=0}^{m}f_i t^i,
\end{eqnarray}
where $m$ is the leading degree of large $t$ behaviour and depends
upon the specific integrand in question.

After applying the three delta functions constraints we see that
taking the residue of $A_1A_2A_3$ at a particular pole, $t=t_0$,
removes any remaining dependence upon the loop momentum. Hence we can
write
\begin{eqnarray}
&&\hspace*{-0.8cm}\int
  \!d^{4}l\prod_{i=0}^2\delta(l_i^2)\frac{\Res_{t=t_0}
  A_1A_2A_3}{t-t_0}\sim\lim_{t\rightarrow
  t_0}\left[(t-t_0)A_1A_2A_3\right]\!\int
  \!d^{4}l\prod_{i=0}^2\delta(l_i^2)\frac{1}{t-t_0}.\label{eq:doing_the_res}
\end{eqnarray}
Where on the right hand side of this we understand the integral, $\int
d^4l$, as over the parameterised form of $l$ in terms of $t$ and the
three other degrees of freedom. In the cut integrand the only source
of poles in $t$ is from propagator terms of the type
$1/(l-P)^2$. Generally each such propagator, when on-shell, contains
two poles due to the quadratic nature, in $t$, of
\eqn{eq:quadratic_sol}. If we label these solutions $t_{\pm}$ then we
can write a triple-cut scalar box in terms of these poles as
\begin{eqnarray}
\int d^{4}l\prod_{i=0}^2\delta(l_i^2)\frac{1}{(l-P)^2}\sim
\frac{1}{t_+-t_-}\left(\int
d^{4}l\prod_{i=0}^2\delta(l_i^2)\frac{1}{t-t_+}-\int
d^{4}l\prod_{i=0}^2\delta(l_i^2)\frac{1}{t-t_-}\right).
\end{eqnarray}
From comparing this to \eqn{eq:doing_the_res} we see that all residue
terms of \eqn{eq:on_shell_split} simply correspond to pieces of
triple-cut scalar box functions multiplied by various coefficients.

Therefore we can associate all residue terms with scalar boxes,
meaning that our triple cut amplitude can be written simply as
\begin{eqnarray}
(2\pi)^3\int d^4l \prod_{i=0}^2\delta(l_i^2)A_1A_2A_3=(2\pi)^3\int dt
J_t\left(\sum_{i=0}^m f_i t^i\right)+\sum_{\textrm{boxes }\{l\}} d_l
D^{\rm cut}_0.\label{eq:trp_cut_almost_1}
\end{eqnarray}
This is a sum over the set $\{l\}$ of possible cut scalar boxes,
$D^{\rm cut}_0$, and their associated coefficients, $d_l$, along with
a power series in positive powers of $t$.  In
\eqn{eq:trp_cut_almost_1} we have integrated over the three delta
functions after performing the integral transformation from $l^{\mu}$
to $t$, the Jacobian of which, and any additional factors picked up
from the integration is contained in the factor $J_t$.  The limit $m$
of the summation is the maximum power of $t$ appearing in the
integrand, which in turn is the maximum power of $l$ appearing in the
numerator of the integrand. In general for renormalisable theories,
such as QCD amplitudes, $m\leq3$.

We must now turn our attention to answering the question of what do
the remaining terms correspond to? To do this we need to understand
the behaviour of the integrals over positive powers of $t$. There is a
freedom in our choice of the parameterisation of the cut-loop
momentum. This freedom extends, as we will prove in
section~\ref{section:the_triple_cut_mom_param}, to choosing a
parametrisation where the integrals over all positive powers of $t$
vanish. Doing this then reduces the cut integrand to
\begin{eqnarray}
(2\pi)^3\int d^4l \prod_{i=0}^2\delta(l_i^2)A_1A_2A_3=(2\pi)^3f_0\int dt
J_t+\sum_{\textrm{boxes }\{l\}} d_l D^{\rm cut}_0.
\end{eqnarray}
The remaining integral is now simply that of a triple-cut scalar
triangle, multiplied by the coefficient $f_0$.  For the triple-cut
scalar triangle integral, $C^{\rm cut}_0(K^2_i,K^2_j)$, given by
$-(2\pi)^3\int dt J_t$, the triple cut form of \eqn{eq:tri_int_def},
we find that its corresponding coefficient is given simply by
\begin{eqnarray}
c_{ij}=-\left[\textrm{Inf}_{t} A_1A_2A_3\right](t)\Big|_{t=0},\label{eq:scalar_triangle_MI}
\end{eqnarray}
which is just the first term in the series expansion in $t$ of the
cut-integrand at infinity.

The simplicity of this result relies crucially upon two facts. The
first is that on the triple cut the integral is sufficiently simple
that it can be decomposed into either a triangle contribution or a box
contribution. This is important as it allows us to easily distinguish
between the two types of term. As an example consider a linear box
which contains a numerator factor constructed such that it vanishes at
the pole contained in the denominator, but without being proportional
to the denominator itself. To which basis integral does this term
contribute to? In the simplest case such a term would look like
\begin{eqnarray}
\int d^4l \prod_{i=0}^2\delta(l_i^2)\frac{\Bmp{lW}}{\Bmp{lP}}=\int d^4l \prod_{i=0}^2\delta(l_i^2)\frac{\Bmp{aW}(t-t_0)}{\Bmp{aP}(t-t_0)}=\frac{\Bmp{aW}}{\Bmp{aP}}\int d^4l \prod_{i=0}^2\delta(l_i^2),\nonumber
\end{eqnarray}
and hence must contribute entirely to the triangle integral, it
contains no box terms. Here we have chosen a simplified loop momentum
parameterisation in terms of two basis spinors $|a^+\rangle$ and
$|\overline{a}^+\rangle$ such that
$\Bmp{lP}=t\Bmp{aP}+\Bmp{\overline{a}P}$. This then contains a pole in
$t$ at $t_0=-\Bmp{\overline{a}P}/\Bmp{aP}$ and we have chosen the
spinor $|W^+\rangle$ such that $\Bmp{\overline{a}W}=-t_0\Bmp{aW}$.

The second crucial fact is the \textit{vanishing} of the other
integrals over $t$ so that the complete scalar triangle integral is
given by only the remaining integral over $t^0$. Hence the coefficient
is given by a single term. Furthermore, the use of a complex loop
momentum also means that we can apply this formalism to the extraction
of scalar coefficients corresponding to one- and two-mass triangles as
well as three-mass triangles. As discussed above for the case of box
coefficients, this is a result of the possibility of a well-defined
three-point vertex when using complex momentum, enabling in these
cases the construction of non-vanishing cut integrands.

\subsection{The momentum parameterisation}
\label{section:the_triple_cut_mom_param}

\begin{figure}[ht]
\centerline{\epsfxsize 2.6 truein\epsfbox{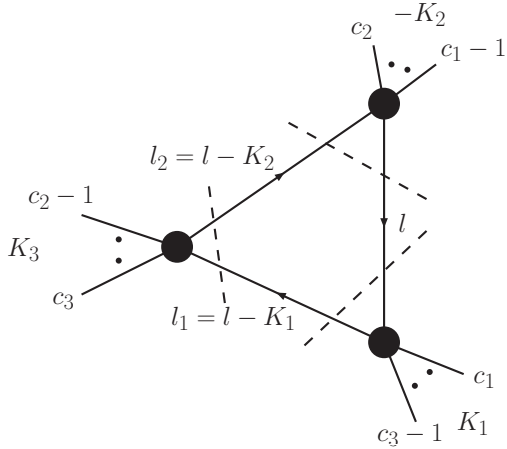}}
\caption{ The triple cut used to compute the  scalar triangle
coefficient of $C_0(K_{1}^2,K^2_{2})$.  }
\label{figure:3_mass_triple_cut}
\end{figure}

We wish to compute the coefficient of the scalar triangle singled out
by the triple cut given in figure \ref{figure:3_mass_triple_cut}.
The cut integral when written in terms of tree amplitudes is
\begin{eqnarray}
&&\hspace*{-0.8cm}(2\pi)^3\int d^4l\!\prod_{i=0}^2\delta(l_i^2)
\At_{c_3-c_1+2}(-l,c_1,\ldots,(c_3-1),l_1)
\At_{c_2-c_3+2}(-l_1,c_3,\ldots,(c_2-1),l_2)
\nonumber \\
&&\hphantom{\int d^4l\!\prod_{i=0}^2\delta(l_i^2)}\times\At_{n-c_2+c_1+2}(-l_2,c_2,\ldots,(c_1-1),l),
\end{eqnarray}
with $l_1=l-K_1=l-K_{c_1\ldots c_3-1}$ and $l_2=l-K_2=l+K_{c_2\ldots
c_1-1}$, so that $K_1=K_{c_1\ldots c_3-1}$ and $K_2=-K_{c_2\ldots
c_1-1}$.

Our first step will be to find a parameterisation of $l$ in terms of
the single free integral parameter remaining after satisfying all
three of the cut delta functions constraints,
\begin{eqnarray}
l^2=0,\;\;\;\;\;l_1^2=(l-K_1)^2=0,\;\;\;{\rm and }\;\;\;\; l_2^2=(l-K_2)^2=0.\label{eq:l_const}
\end{eqnarray}
Each of the three legs can be massive or massless. We will deal with
the general case of three massive legs explicitly here. The cases with
massless legs are then easily found by setting the relevant mass in
the parameterisation to zero. We will find it very convenient to
express $l^{\mu}$ in terms of a basis of momentum identical to the
momenta $l_1$ and $l_2$ used by Ossola, Papadopoulos and
Pittau~\cite{Ossola:2006us}. We will write these momenta in the
suggestive notation $\Kf1$ and $\Kf2$ and define them via
\begin{eqnarray}
K_1^{\flat,\mu}&=&K^{\mu}_1-\frac{S_1}{\gamma}K^{\flat,\mu}_2,\nonumber\\
K_2^{\flat,\mu}&=&K_2^{\mu}-\frac{S_2}{\gamma}K^{\flat,\mu}_1,
\end{eqnarray}
with $\gamma=\Bmp{K_1^{\flat,-}|\s K_2^{\flat}|K_1^{\flat,-}}
\equiv\Bmp{K_2^{\flat,-}|\s K_1^{\flat}|K_2^{\flat,-}}$ and
$S_i=K^2_i$.  Each momentum $\Kf1$, $\Kf2$ is the massless projection
of one of the massive legs in the direction of the other masslessly
projected leg. A more practical definition of $\Kf1$ and $\Kf2$, in
terms of the external momenta alone, can be found by solving the above
equations for $\Kf1$ and $\Kf2$, so that in terms of $S_1$, $S_2$,
$K_1^{\mu}$ and $K_2^{\mu}$ we have
\begin{eqnarray}
K_1^{\flat,\mu}=\frac{K^{\mu}_1-(S_1/\gamma)K_2^{\mu}}{1-(S_1S_2/\gamma^2)},
\;\;\;\;\;K_2^{\flat,\mu}=\frac{K_2^{\mu}-(S_2/\gamma)K^{\mu}_1}{1-(S_1S_2/\gamma^2)}.\label{eq:def_gamma_pm}
\end{eqnarray}
In addition $\gamma$ can be expressed in terms of the external
momenta,
\begin{eqnarray}
\gamma_{\pm}=(K_1\cdot K_2)\pm\sqrt{\Delta},\;\;\;\;\;\;\Delta=(K_1\cdot K_2)^2-K_1^2K_2^2.\label{eq:def_delta}
\end{eqnarray}
When using \eqn{eq:scalar_triangle_MI} we must average over the number
of solutions of $\gamma$.  In the three-mass case there are a pair of
solutions. For the one- and two-mass cases, when either $K_1^2=0$ or
$K_2^2=0$, then there is only a single solution.

After satisfying the three constraints given by \eqn{eq:l_const} we
write the spinor components of $l^{\mu}$ in terms of our basis $\Kf1$
and $\Kf2$ as
\begin{eqnarray}
\langle l^-|&=&t\langle K_{1}^{\flat,-}|+\alpha_{01}\langle K_{2}^{\flat,-}|,
\nonumber\\
\langle
l^+|&=&\frac{\alpha_{02}}{t}\langle K_{1}^{\flat,+}|+\langle K_{2}^{\flat,+}|,
\label{eq:three_mass_mom_param}
\end{eqnarray}
where
\begin{eqnarray}
\alpha_{01}=\frac{S_1\left(\gamma-S_2\right)}{\left(\gamma^2-S_1S_2\right)},\;\;\;\;\;\;\alpha_{02}=\frac{S_2\left(\gamma-S_1\right)}{\left(\gamma^2-S_1S_2\right)}.
\end{eqnarray}
Written as a four-vector, $l^{\mu}$ is given by
\begin{eqnarray}
l^{\mu}=\alpha_{02}K_1^{\flat,\mu}+\alpha_{01} K_2^{\flat,\mu}+\frac{t}{2}\Bmp{K^{\flat,-}_1|\gamma^{\mu}|K^{\flat,-}_2}
+\frac{\alpha_{01}\alpha_{02}}{2t}\Bmp{K^{\flat,-}_2|\gamma^{\mu}|K^{\flat,-}_1}.\label{eq:sol_for_l_in_3_mass}
\end{eqnarray}

We can also use momentum conservation to write component forms for the
other two cut momenta $l_i$ with $i=1,2$,
\begin{eqnarray}
\langle l_i^-|
&=&t\langle K_{1}^{\flat,-}|+\alpha_{i1}\langle K_{2}^{\flat,-}|,
\nonumber\\
\langle l_i^+|
&=&\frac{\alpha_{i2}}{t}\langle K_{1}^{\flat,+}|+\langle K_{2}^{\flat,+}|,
\end{eqnarray}
where the $\alpha_{ij}$ are given in
Appendix~\ref{appendix:the_triangle_param}.

A final point is that after having integrated over the three delta
function constraints and performed the change of variables to the
momentum parameterisation of \eqn{eq:three_mass_mom_param} we have the
factor $J_t=1/(t\gamma)$ contained in \eqn{eq:trp_cut_almost_1}. We
always associate this factor with the scalar triangle integral and so
its explicit form does not play a role in our formalism.

\subsection{Vanishing integrals}
\label{section:vanishing_ints}

As we have remarked previously, the simplicity of the method outlined
here rests crucially upon the properties of the momentum
parameterisation we have used. The key feature is the
\textit{vanishing} of the integrals over $t$. It can easily be shown
that within our chosen momentum parameterisation, of
section~\ref{section:the_triple_cut_mom_param}, any integral of a
positive or negative power of $t$ vanishes. Following an argument very
similar to that used by Ossola, Papadopoulos and
Pittau~\cite{Ossola:2006us} we use $\Bmp{\Kfpm1|\s K_1|\Kfpm2}=0$,
$\Bmp{\Kfpm1|\s K_2|\Kfpm2}=0$ and
$\Bmp{\Kfpm1|\gamma^{\mu}|\Kfpm2}\Bmp{\Kfpm1|\gamma_{\mu}|\Kfpm2}=0$,
to show that
\begin{eqnarray}
&&\int d^4l \frac{\Bmp{\Kfm1|\s l|\Kfm2}^n}{l^2l_1^2l_2^2}=0
\;\;\;\Rightarrow\;\;\;\int dt J_t \frac{1}{t^n}=0 \;\;\textrm{ for } n\geq1,
\nonumber\\
&&\int d^4l \frac{\Bmp{\Kfm2|\s
    l|\Kfm1}^n}{l^2l_1^2l_2^2}=0 \;\;\;\Rightarrow\;\;\;\int dt J_t t^n =0 \;\;\textrm{ for } n\geq1.
\end{eqnarray}
The vanishing of these terms then leads directly to our general
procedure, encapsulated in \eqn{eq:scalar_triangle_MI}, which is to
simply express the triple cut of the desired scalar triangle in the
momentum parameterisation given by \eqn{eq:three_mass_mom_param} and
then take the $t^0$ component of a series expansion in $t$ around
infinity.


\section{Two-particle cuts and scalar bubble coefficients}
\label{section:double_cuts_and_bubble_coeffs}

In the same spirit as the triangle case we now wish to extract the
coefficients of scalar bubble terms using, in this case, a
two-particle cut. Now a two-particle cut will contain in addition to
our desired scalar bubble both scalar boxes and triangles, all of
which need to be disentangled. What we will find, though, is that
naively applying the technique as given for the scalar triangle
coefficients will not give us the complete scalar bubble contribution.

The reason for this is straightforward to see. A two-particle cut
places only two constraints on the loop momentum and so we can
parameterise it in terms of two free variables, which we will label
$t$ and $y$. Consider rewriting the cut integrand in a partial
fraction decomposition in terms of $y$. Schematically, therefore, the
two-particle cut of the legs $l^2$ and $(l-K_1)^2$ can be written as
\begin{eqnarray}
(2\pi)^2\!\!\int \!d^4l \prod_{i=0}^1\delta(l_i^2)A_1A_2
\!=\!(2\pi)^2\!\!\int\!\! dt dy J_{t,y}\!\left( \left[\textrm{Inf}_{y} A_1A_2\right](y)
+\sum_{\textrm{poles }\{j\}} \frac{\Res_{y=y_j}
A_1A_2}{y-y_j}\right),\label{eq:bubble_gen_cut}
\end{eqnarray}
where again $\{j\}$ is the sum over all possible poles, this time in
$y$, and $J_{t,y}$ contains any terms from the change into the
parameterisation of $y$ and $t$ as well as any pieces picked up by
integrating over the two delta functions. So far this seems to be
similar to the triangle case, but with the residue terms now
corresponding to triangles as well as boxes. As we have two parameters
though we can consider a further partial fraction decomposition, this
time with $t$, giving
\begin{eqnarray}
&&\hspace*{-0.8cm}(2\pi)^2\int d^4l \prod_{i=0}^1\delta(l_i^2)A_1A_2=
\nonumber\\
&&(2\pi)^2\int dt dy J_{t,y}\left( \left[\textrm{Inf}_{t}\left[\textrm{Inf}_{y} A_1A_2\right](y)\right](t)
+\left[\textrm{Inf}_t\left(\sum_{\textrm{poles }\{j\}} \frac{\Res_{y=y_j}
A_1A_2}{y-y_j}\right)\right](t)\right.
\nonumber\\
&&\hphantom{(2\pi)^3\int}+\left.\sum_{\textrm{poles }\{l\}} \frac{\Res_{t=t_l}
\left[\textrm{Inf}_{y}A_1A_2\right](y)}{t-t_l}
+\sum_{\textrm{poles }\{j\},\{l\}}\frac{\Res_{t=t_l}\left[\frac{\Res_{y=y_j}A_1A_2}{y-y_j}\right]}{t-t_l}
\right),\label{eq:bubble_gen_cut_full}
\end{eqnarray}
where here $\{l\}$ is the sum over all possible poles in $t$.  The
general dependence of the cut integral momentum, $l^{\mu}$, on the
free integral parameters $t$ and $y$ can be written in terms of null
four-vectors $a_i^{\mu}$ with $i=0,1,2,3,4$ such that $l^2=0$. An
explicit form for these will be presented in
section~\ref{section:bubble_cut_mom_param}.  We then define $l^{\mu}$
by
\begin{eqnarray}
l^{\mu}= \frac{y^2}{t}a_0^{\mu}+\frac{y}{t}a_1^{\mu}+y a_2^{\mu}+t
a_3^{\mu}+a_4^{\mu}.
\end{eqnarray}
Again residues of pole terms will correspond to the solutions of
$(l-P)^2=0$ and hence it is straightforward to see that the final term
of \eqn{eq:bubble_gen_cut_full}, containing the sum of residues in
both $y$ and $t$, has both of these free parameters fixed. Any such
terms must contain at least one propagator pole. Also the numerator
will be independent of any integration variables, as both $y$ and $t$
are fixed. Thus all such terms will correspond to purely scalar
triangle and scalar box terms. Looking at the second and third terms
of \eqn{eq:bubble_gen_cut_full} we might also, at least initially,
want to associate these terms with contributions to scalar triangle
terms only and hence naively conclude that only the first term of
\eqn{eq:bubble_gen_cut_full} contributes to the scalar bubble
coefficient. This assumption though would be wrong.

The crucial difference between the single residue terms of
\eqn{eq:bubble_gen_cut_full} and those of \eqn{eq:on_shell_split} is
the parameterisation of the loop momentum which is being used. Taking
the residue of a pole term at a particular point $y$ freezes $y$ such
that we force a particular momentum parameterisation upon these
triple-cut terms. Importantly, in general this particular forced
momentum parameterisation is such that the integrals over $t$ in the
second and third terms of \eqn{eq:bubble_gen_cut_full} now no longer
vanish.

If only scalar triangle contributions came from the integrals over $t$
then this would not be an issue; we could just discard these terms as
not relevant for the extraction of our bubble coefficient. What we
find though, through a simple application of Passarino-Veltman
reduction techniques, is that these integrals contain scalar bubble
contributions, $B_0$, with coefficients $b$,
\begin{eqnarray}
 \int dt J'_t t^n=b\, B_0+c \,C_0,\label{eq:non_vanishing_tri}
\end{eqnarray}
where $J'_t$ is the relevant Jacobian for this parameterisation of the
loop momentum and $c$ is the coefficient corresponding to the scalar
triangle contribution, $C_0$. We cannot therefore simply discard the
residue pieces of \eqn{eq:bubble_gen_cut_full}, as we could in the
triangle case, if we want to derive the full scalar bubble
coefficient. Furthermore, there is an additional complication. We will
see that the integrals over powers of $y$ contained in the first term
of \eqn{eq:bubble_gen_cut_full} also do not vanish in general and
hence must also be taken into account.

There is a limit to the maximum positive powers of $y$ and $t$ that
appear in the rewritten partial-fractioned decomposition of the
integral. For renormalisable theories, such as QCD, up to three powers
of $t$ appear for triangle coefficients and up to four powers of $y$
for bubble coefficients.  Therefore the power series in $y$ and $t$ of
the \Inf{} operators will always terminate at these fixed points. It
is then straightforward, as we will discuss in
section~\ref{section:non_vanishing_ints} and
section~\ref{section:performing_the_shift_in_y_and_t}, to derive the
general form for all possible non-vanishing contributing integrals,
over powers of $y$ and $t$, in terms of their scalar bubble
contributions.

Calculation of the scalar bubble coefficient therefore requires a two
stage process. First take the $\textrm{Inf}_{y}$ and
$\textrm{Inf}_{t}$ pieces of the cut integrand and replace any
integrals over $y$ with their known \textit{general} forms, as we
shall see integrals proportional to $t$ will vanish. Secondly compute
all possible triple cuts that could be generated by applying a third
cut to the two-particle cut we are considering. To these terms then
apply, not the parameterisation we used in
section~\ref{eq:triple_cuts_and_scalar_triangles}, but the
parameterisation forced upon us by taking the residues of the poles in
$y$, which we will derive in
section~\ref{section:forced_triple_mom}. This is equivalent to
calculating all the contributions from the residues of the partial
fraction decomposed cut integrand of
\eqn{eq:bubble_gen_cut_full}. Within these terms we then replace any
integrals of powers of $t$ with their known \textit{general}
forms. Finally we sum all the contributing pieces together to get the
full scalar bubble contribution and hence its coefficient. Our final
result for assembling the bubble coefficient is then given by
\eqn{eq:scalar_coeff_result}.

\subsection{The momentum parameterisation for the two-particle cut}
\label{section:bubble_cut_mom_param}

\begin{figure}[ht]
\centerline{\epsfxsize 3 truein\epsfbox{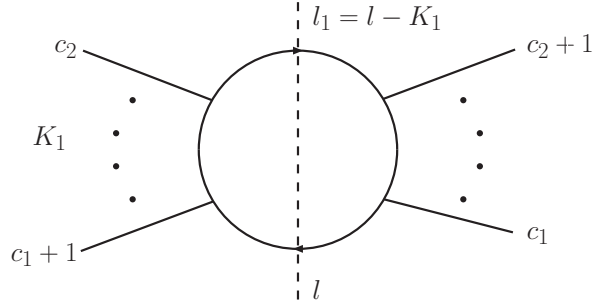}}
\caption{ The two-particle cut for computing the scalar bubble
coefficient of $B_0(K^2_{1})$.  }
\label{figure:double_cut}
\end{figure}

We want to extract the scalar bubble coefficient obtainable from the
application of the two-particle cut given in figure
\ref{figure:double_cut}.
This two-particle cut can be expressed in terms of tree amplitudes as
\begin{eqnarray}
&&\hspace*{-0.8cm}(2\pi)^2\!\!\int
d^4l\prod_{i=0}^1\delta(l_i^2)
\At_{c_2-c_1+2}(-l,(c_1+1),\ldots,c_2,l_1)
\At_{n-c_2+c_1+2}(-l_1,(c_2+1),\ldots,c_1,l),
\end{eqnarray}
with $l_1=l-K_{c_1+1\ldots c_2}=l-K_{1}$.

A bubble can be classified entirely in terms of the momentum of one of
its two legs, which we label $K_1$, and so we will find it useful to
express the cut loop momentum $l$ in terms of the pair of massless
momenta $\Kf1$ and $\chi$ defined via
\begin{eqnarray}
K_1^{\flat,\mu}=K_1^{\mu}-\frac{S_1}{\gamma}\chi^{\mu},
\end{eqnarray}
here $\gamma=\Bmp{\chi^\pm|\s K_1|\chi^{\pm}}\equiv\Bmp{\chi^\pm|\s
{\Kf1}|\chi^{\pm}}$. The arbitrary vector $\chi$ can be chosen
independently for each bubble coefficient as a result of the
independence of the choice of basis representation for the cut
momentum. In the two-particle cut case we have only two momentum
constraints
\begin{eqnarray}
l^2=0,\;\;\;{\rm and }\;\;\;l_1^2=(l-K_1)^2=0,
\end{eqnarray}
and so we have two free parameters which we will label $y$ and
$t$. The loop momentum can then be expressed in terms of spinor
components as
\begin{eqnarray}
\langle l^-|&=&t\langle
K_{1}^{\flat,-}|+\frac{S_1}{\gamma}\left(1-y\right)\langle
\chi^{-}|,
\nonumber\\
\langle l^+|&=&\frac{y}{t}\langle K_{1}^{\flat,+}|+\langle \chi^{+}|.\label{eq:dble_cut_mom_def}
\end{eqnarray}
Written as a four-vector $l^{\mu}$ is
\begin{eqnarray}
l^{\mu}&=&y K_1^{\flat,\mu}+\frac{S_1}{\gamma}(1-y)\chi^{\mu}
+\frac{t}{2}
\Bmp{\Kfm1|\gamma^{\mu}|\chi^-}
+\frac{S_1}{2\gamma}\frac{y}{t}(1-y)\Bmp{\chi^-|\gamma^{\mu}|\Kfm1}.\label{eq:bubble_cut_4_vector}
\end{eqnarray}

We can also use momentum conservation to write a component form for
the other cut momentum. We have
\begin{eqnarray}
\langle l_1^-|
&=&\langle K_{1}^{\flat,-}|-\frac{S_1}{\gamma}\frac{y}{t}\langle \chi^{-}|,
\nonumber\\
\langle l_1^+|
&=&\left(y-1\right)\langle K_{1}^{\flat,+}|+t\langle \chi^{+}|.
\end{eqnarray}

Furthermore after rewriting the integral in this cut-momentum
parameterisation and integrating over the two delta function
constraints we find the following simple result for the constant
$J_{t,y}$ contained in \eqn{eq:bubble_gen_cut}, namely $J_{t,y}=1$.

\subsection{Non-vanishing integrals}
\label{section:performing_the_shift_in_y_and_t}

In the case of the scalar triangles of
section~\ref{section:vanishing_ints} crucial simplifications occurred
as a result of our chosen cut momentum parameterisation. Any integral
over a power of $t$ vanished, leaving only a single contribution
corresponding to the desired coefficient. For the scalar bubble
coefficient things are not quite as simple.

We can use $\Bmp{\Kfpm1|\s K_1|\chi^{\pm}}=0$ as well as
$\Bmp{\Kfpm1|\gamma^{\mu}|\chi^{\pm}}\Bmp{\Kfpm1|\gamma_{\mu}|\chi^{\pm}}=0$
to show that
\begin{eqnarray}
\int d^4l\frac{\Bmp{\chi^-|\s l|\Kfm1}^n}{l^2l_1^2}&=&0\;\;\;\Rightarrow\;\;\;
\int dtdy \;t^n=0,
\nonumber\\
\int d^4l\frac{\Bmp{\Kfm1|\s l|\chi^-}^n}{l^2l_1^2}&=&0\;\;\;\Rightarrow\;\;\;\int
dtdy 
\left(\frac{y}{t}\right)^n(1-y)^n =0.
\end{eqnarray}
Hence the integrals over all positive and negative powers of $t$
vanish,
\begin{eqnarray}
\int dt dy \;t^n=0 \;\;\textrm{ for }n\neq0.
\end{eqnarray}

Integrals over positive powers of $y$, contained within the double
\Inf{} piece of the first term of \eqn{eq:bubble_gen_cut_full}, will
\textit{not} vanish. These integrals are straightforwardly derivable
with the aid of identities involving the four vector
$n^{\mu}=K_1^{\flat,\mu}-(S_1/\gamma)\chi^{\mu}$ which satisfies the
constraints $(K_1\cdot n)=0$ and $n^2=-S_1$. It is then possible to
show the following relations in $D=4$ dimensions, and remembering that
$J_{t,y}=1$,
\begin{eqnarray}
&&\int d^4l\frac{(l\cdot n)^{2m-1}}{l^2l_1^2}=0\;\;\;\Rightarrow \;\;\;\int dt dy\left(\frac{1}{2}-y\right)^{2m-1}=0,
\nonumber\\
&&\int d^4l
\frac{(l\cdot n)^{2m}}{l^2l_1^2}=S_1^{2m}B_{\mathcal{P}\mathcal{V}}^m\;\;\;\Rightarrow\;\;\; \int dt dy\left(\frac{1}{2}-y\right)^{2m}=S_1^{2m}\tilde{B}^m_{\mathcal{P}\mathcal{V}},
\nonumber\\
&&\int d^4l
\frac{(l\cdot K_1)^{2m}}{l^2l_1^2}=(2m+1)S_1^{2m}B^m_{\mathcal{P}\mathcal{V}}\;\;\;\Rightarrow\;\;\; \frac{1}{2^{2m}}\int dt dy=(2m+1)S_1^{2m}\tilde{B}^m_{\mathcal{P}\mathcal{V}},
\end{eqnarray}
where $B^m_{\mathcal{P}\mathcal{V}}$ and
$\tilde{B}^m_{\mathcal{P}\mathcal{V}}$ are Passarino-Veltman reduction
coefficients, the explicit forms of which are not needed. Solving
these equations for the integral of $y^m$ leads to the result
\begin{eqnarray}
\int dt dy \;y^m=\frac{1}{m+1}\int dt dy \;\;\textrm{   for }m\geq 0.
\end{eqnarray}

Contributions to our desired scalar bubble coefficient from the double
\Inf{} piece of \eqn{eq:bubble_gen_cut_full} therefore come not only
from the single constant $t^0y^0$ term but also from terms
proportional to integrals of $t^0y^m$. This is not the end of the
story. As described above, there can be further contributions from the
second and third residue terms generated in the decomposition of
\eqn{eq:bubble_gen_cut_full}. We could proceed from the cut integrand
to explicitly calculate these residue terms. However as we will shall
see, a more straightforward approach is to derive these terms by
relating them to triple cuts.

\subsection{The momentum parameterisation for triple cut contributions}
\label{section:forced_triple_mom}

We wish to relate the contributions to the bubble coefficient of the
residue pieces, separated in the decomposition of
\eqn{eq:bubble_gen_cut_full}, to triple cuts in a specific basis of
the cut-loop momentum. To find this basis we will apply the additional
constraint
\begin{eqnarray}
(l+K_2)^2=0,\label{eq:extra_constr}
\end{eqnarray}
to the two-particle cut momentum of
section~\ref{section:bubble_cut_mom_param}. Note that here we label
the ``$K_2$'' leg as $K_2$ in contrast to $(-K_2)$ as we did in the
triangle coefficient case of
section~\ref{section:the_triple_cut_mom_param}.  This constraint
corresponds to the application of an additional cut which would appear
as $\delta((l+K_2)^2)$ inside the integral. This additional
constraint, applied to the starting point of the two-particle cut loop
momentum, forces us to use $\Kf1$ and $\chi$ as the momentum basis
vectors of $l$. Importantly, this differs from the basis choice for the
triple cut momenta developed in
section~\ref{section:the_triple_cut_mom_param}, which leads to the
differing behaviour of these triple-cut contributions.

The presence of $y$ in both $\langle l^-|$ and $\langle l^+|$ directs
us for reasons of efficiency to choose to use \eqn{eq:extra_constr} to
first constrain $y$, leaving $t$ free. Looking at
\eqn{eq:bubble_cut_4_vector} we see that as $l^{\mu}$ is quadratic in
$y$ then there are two solutions to this constraint, $y_{\pm}$, which
are given by
\begin{eqnarray}
y_{\pm}&\!=&\!\frac{1}{2S_1\Bmp{\chi^-|\s
    K_2|\Kfm1}}\Bigg(\!\!\left(\gamma\Bmp{\Kfm1|\s K_2|\Kfm1} -S_1\Bmp{\chi^-|\s
    K_2|\chi^-}\right)t
+S_1\Bmp{\chi^-|\s K_2|\Kfm1}
\nonumber\\
&&\!\!\pm\sqrt{\!\left(\!S_1\Bmp{\chi^-|\s K_2|\Kfm1}+2t \gamma \left(K_1\cdot
    K_2\right)\right)^2-4 S_1S_2 \gamma t \left(t\gamma-\Bmp{\chi^-|\s
    K_2|\Kfm1}\right)}\Bigg).
\label{eq:def_of_y_for_triple_cut}
\end{eqnarray}
On substituting these two solutions into the two-particle cut momentum
of \eqn{eq:dble_cut_mom_def} we obtain our desired triple-cut momentum
parameterisation.

Our final step is then to relate the triple-cut integrals defined in
this basis to the residue terms of
\eqn{eq:bubble_gen_cut_full}. Rewriting the triple cut integral after
the change of momentum parameterisation and integrating over all but
the third delta function gives the general form
\begin{eqnarray}
&&\hspace*{-0.8cm}(2\pi)^3\int dt dy\;J'_t
\left(\delta(y-y_{+})+\delta(y-y_{-})\right)\mathcal{M}(y,t),
\end{eqnarray}
where $\mathcal{M}(y,t)$ is a general cut integrand and
\begin{eqnarray}
J'_t=\frac{1}{\sqrt{\left(S_1\Bmp{\chi^-|\s K_2|\Kfm1}+2t \gamma \left(K_1\cdot
    K_2\right)\right)^2-4 S_1S_2 \gamma t \left(t\gamma-\Bmp{\chi^-|\s
    K_2|\Kfm1}\right)}}.\label{eq:def_Jtp_for_forced_param}
\end{eqnarray}
Upon examination of a general residue term we find that it corresponds
to an integral of the form
\begin{eqnarray}
&&\hspace*{-0.8cm}(2\pi)^2i\!\!\int \!dt dy J_{t,y}\Res_{y=y_{\pm}}\frac{\mathcal{M}(y,t)}{(l+K_2)^2}\equiv-\frac{(2\pi)^3}{2}\!\!\int \!dt dy\;J'_t
\left(\delta(y-y_{+})+\delta(y-y_{-})\right)\mathcal{M}(y,t),
\end{eqnarray}
and  hence that residue  contributions are  given, up  to a  factor of
$(-1/2)$, by the triple cut.

This result applies equally when $S_2=K_2^2=0$, corresponding to a one
or two-mass triangle, when the appropriate scale is set to zero in
\eqn{eq:def_of_y_for_triple_cut} and
\eqn{eq:def_Jtp_for_forced_param}. The momentum parameterisation in
this simplified case is contained in
Appendix~\ref{appendix:forced_triple_cut_S_2_0}.

\subsection{More non-vanishing integrals and bubble coefficients}
\label{section:non_vanishing_ints}

There is a direct correspondence between a triple cut contribution and
a residue contribution. The sum of all possible triple cuts, which
contain the original two-particle cut, will therefore correspond to
the sum of all residue terms. We must now examine how such terms
contribute to the bubble coefficient itself.

Unlike for the case of triple cut integrands as parameterised in
section~\ref{section:the_triple_cut_mom_param} we will find that there
are contributions, specifically in this case bubble coefficient
contributions, coming from the integrals over $t$. To see this let us
investigate the integrals over $t$ in more detail. As an example
consider extracting the scalar bubble term coming from a two-mass
linear triangle (with the massless leg $K_2$ so that $S_2=0$). We
would start from a two-particle cut which, after decomposing as
\eqn{eq:bubble_gen_cut}, would give
\begin{eqnarray}
&&\hspace*{-0.8cm}(2\pi)^2\int d^4l
    \prod_{i=0}^1\delta(l_i^2)\frac{\Bmp{K_2^-|\s l|a^-}}{(l+K_2)^2}
\label{eq:non_vanishing_buble_shift_1}\\
&=&(2\pi)^2\int d^4l
    \prod_{i=0}^1\delta(l_i^2)\frac{[la]}{[lK_2]}
=(2\pi)^2\int \! dt dyJ'_t\!\left(\!\frac{[\Kf1 a]}{[\Kf1
    K_2]}+\!\!\frac{\Res_{y= -t \frac{[\chi K_2]}{[\Kf1
    K_2]}}\!\left(\frac{y[\Kf1a]+t[\chi a]}{y+t\frac{[\chi
    K_2]}{[\Kf1 K_2]}}\right)}{[\Kf1 K_2]\left(\frac{y}{t}[\Kf1
    K_2]+[\chi K_2]\right)}\!\right).\nonumber
\end{eqnarray}
The first term of this is clearly not the complete coefficient, and so
we need to obtain the bubble contribution contained within the second
term. Consider reconstructing this term using a triple cut with the
cut loop momentum parameterised in a form given by setting $y$ equal
to its value at the residue of the pole of this second term. This
triple cut term is given by
\begin{eqnarray}
&&\hspace*{-0.8cm}-\frac{(2\pi)^3i}{2}\!\int\! d^4l\!
  \prod_{i=0}^2\delta(l_i^2)\Bmp{K_2^-|\s  l|a^-}
\nonumber\\
&&=-\frac{(2\pi)^3i}{2}\!\int dt J'_t\frac{[\chi \Kf1][K_2a]}{[\Kf1
    K_2]^2}\left(\!\!\Bmp{K_2^-|\s
    K_1|K_2^-}t\!+\!\frac{S_1}{\gamma}\Bmp{\chi^-|\s K_2|\Kfm1}\!\right),
\label{eq:non_vanish_int_examp_1}
\end{eqnarray}
where we have used the parameterisation of $l^{\mu}$ given by
\eqn{eq:forced_triple_yminus_sol_l} and added an extra overall factor
of $i$ which would come from the additional tree amplitude in a triple
cut.

Of this triple cut integrand only the first, $t$ dependent, term can
give anything other than a scalar triangle contribution. To derive the
result of this integral over $t$ we will, as we have done previously,
use our parameterisation of the cut momentum,
\eqn{eq:bubble_cut_4_vector}, to pick out the integral as follows
\begin{eqnarray}
-i\int d^4l\frac{\Bmp{\chi^-|\s l|\Kfm1}}{l^2l_1^2(l+K_2)^2}\equiv (2\pi)^3\gamma\int dt J'_tt.
\end{eqnarray}
Using Passarino-Veltman reduction on the single tensor integral on the
left hand side of this as well as dropping anything but the
contributing bubble integrals of our particular cut leaves us with the
result
\begin{eqnarray}
\int dt J'_t t=\frac{2}{(2\pi)^3}\frac{S_1\Bmp{\chi 
  K_2}[K_2\Kf1]}{\gamma\Bmp{K_2^-|\s K_1|K_2^-}^2}B^{\rm cut}_0(K^2_1),\label{eq:x4_solution_t}
\end{eqnarray}
where $B^{\rm cut}_0(K^2_1)$ is the cut form of the scalar bubble
integral of \eqn{eq:bubble_int}.  This non-vanishing result for the
integral over $t$, in contrast to that of
section~\ref{section:the_triple_cut_mom_param}, is a direct
consequence of the cut momentum parameterisation forced upon us when
taking the residues contained in the two-particle cut integrand with
which we started.

On substituting the result of \eqn{eq:x4_solution_t} into
\eqn{eq:non_vanish_int_examp_1} we find that we can write
\eqn{eq:non_vanishing_buble_shift_1}, using the bubble integral given
in \eqn{eq:bubble_int}, as
\begin{eqnarray}
(2\pi)^2\!\!\int d^4l \prod_{i=0}^1\delta(l_i^2)\frac{\Bmp{K_2^-|\s l|a^-}}{(l+K_2)^2}
\!&=&\!\!\! -i\frac{[\chi \Kf1][K_2a]}{[\Kf1
    K_2]^2}\frac{S_1}{\gamma}\frac{\Bmp{\chi^-|\s K_2|\Kfm1}}{\Bmp{K_2^-|\s
    K_1|K_2^-}}B^{\rm cut}_0(K^2_1)\!+\!i\frac{[\Kf1 a]}{[\Kf1 K_2]}B^{\rm cut}_0(K^2_1)
\nonumber\\
&=&i\frac{\Bmp{K_2^-|\s K_1|a^-}}{\Bmp{K_2^-|\s
    K_1|K_2^-}}B^{\rm cut}_0(K^2_1),\label{eq:non_vanish_int_examp_1a}
\end{eqnarray}
which is the known coefficient of the scalar bubble contained inside
the linear triangle.

Of course, if we had chosen $\chi=K_2$ from the beginning, then the
first term on the left hand side of
\eqn{eq:non_vanishing_buble_shift_1} would have been the complete
bubble coefficient. In general, if we are able to rewrite a
two-particle cut integrand such that each term contains only a single
propagator then we can always choose a different $\chi=\Kf2$,
defined via
\begin{eqnarray}
K^{\flat,\mu}_2=K^{\mu}_2-\frac{S_2}{\Bmp{\Kfm1|\s
K_2|\Kfm1}}K^{\flat,\mu}_1,\label{eq:def_K2flat_for_bub}
\end{eqnarray}
for each term individually such that there are no contributions from
the residue terms. Whether this is both feasible and a more
computationally effective approach than calculating the residue
contributions through the use of triple cuts would depend upon the cut
integrand in question.

In general we will be considering processes which contain terms with
powers of up to $t^3$, so we will need to know these integrals. Again
these can be found using a straightforward application of tensor
reduction techniques. When all three legs in the cut are massive these
integrals over $t$ are given, after dropping an overall factor of
$1/(2\pi)^3$ witch always cancels out of the final coefficient, by
\begin{eqnarray}
\!\!\!\!\!\!T(j)=\!\!\int \!dt J'_t t^j\!=\!\left(\frac{S_1}{\gamma}\right)^j\!\!\frac{\Bmp{\chi^-|\s 
  K_2|\Kfm1}^j(K_1\cdot K_2)^{j-1}}{\Delta^{j}}\!\left(\sum_{l=1}^{j}\mathcal{C}_{jl}\frac{S_2^{l-1}}{(K_1\cdot K_2)^{l-1}}\right)\!B^{\rm cut}_0(K^2_1).\label{eq:the_triangle_t_ints}
\end{eqnarray}
Simply taking the relevant mass to zero gives the forms in the one and
two mass cases. $\Delta$ was previously defined in \eqn{eq:def_delta}
and we have
\begin{eqnarray}
&&\mathcal{C}_{11}=\frac{1}{2},
\nonumber\\
&&\mathcal{C}_{21}=-\frac{3}{8},\;\;\;\mathcal{C}_{22}=-\frac{3}{8},
\nonumber\\
&&\mathcal{C}_{31}=-\frac{1}{12}\frac{\Delta}{(K_1\cdot K_2)^2}+\frac{5}{16},\;\;\;\mathcal{C}_{32}=\frac{5}{8},\;\;\;\mathcal{C}_{33}=\frac{5}{16}.
\end{eqnarray}
Also for later use we define $T(0)=0$.

\subsection{The bubble coefficient}

We have now assembled all the pieces necessary to compute our desired
scalar bubble coefficient, $b_j$, corresponding to the cut scalar
bubble integral $B^{\rm cut}_0(K^2_j)$. It is given in general not as
the coefficient of a single term but by summing together the $t^0y^m$
terms from both the double \Inf{} in $y$ followed by $t$ as well as
residue contributions which we derive by considering all possible
triple cuts contained in the two-particle cut. The coefficient is
given by
\begin{eqnarray}
b_j=-i\left[\textrm{Inf}_t\left[\textrm{Inf}_y A_1A_2\right](y)\right](t)\Big|_{t\rightarrow
0,\;y^m\rightarrow \frac{1}{m+1}}-\frac{1}{2}\sum_{\{\mathcal{C}_{\rm
tri}\}} \left[\textrm{Inf}_t A_1A_2A_3\right](t)\Big|_{t^j\rightarrow T(j)},\label{eq:scalar_coeff_result}
\end{eqnarray}
where $T(j)$ is defined in \eqn{eq:the_triangle_t_ints} and the sum
over the set $\{\mathcal{C}_{\rm tri}\}$ is a sum over all triple cuts
obtainable by cutting one more leg of the two-particle cut integrand
$A_1A_2$.

When computing with \eqn{eq:scalar_coeff_result} there is a freedom in
the choice of $\chi$. A suitable choice of which can simplify the
degree of computation involved in extracting a particular coefficient.
Particular choices of $\chi$ can eliminate the need to calculate the
second term of \eqn{eq:scalar_coeff_result} completely, as discussed
in section~\ref{section:non_vanishing_ints}. We also note that there
are choices of $\chi$ which eliminate the need to evaluate the first
term of \eqn{eq:scalar_coeff_result}, so that the coefficient comes
entirely from the second term of \eqn{eq:scalar_coeff_result} instead.

\section{Applications}
\label{section:examples}

To demonstrate our method we now present the recalculation of some
representative triangle and bubble integral coefficients. We also
discuss checks we have made against other various state-of-the-art
cut-constructable coefficients contained in the literature.

\subsection{Extracting coefficients}

To highlight the application of our procedure to the extraction of
basis integral coefficients we consider deriving the coefficients of
some simple integral functions which commonly appear, for example,
in one-loop Feynman diagrams.

\subsubsection{The triangle coefficient of a linear two-mass triangle}

First we consider deriving the scalar triangle coefficient of a linear
two-mass triangle with massive leg $K_1$, massless leg $K_2$, and $a$
and $b$ arbitrary massless four-vectors not equal to $K_2$. This is given
by the integral
\begin{eqnarray}
-i\int d^4l\frac{\Bmp{a^-|\s l|b^-}}{l^2(l-K_1)^2(l+K_2)^2}.
\end{eqnarray}
Extracting the triangle coefficient requires cutting all three
propagators of the integrand. We do this here by simply removing the
``cut'' propagator as we are interested only in the integrand. This
leaves only
\begin{eqnarray}
\Bmp{a^-|\s l|b^-}.\label{eq:exampl_1_bit_1}
\end{eqnarray} 
Rewriting this integrand in terms of the parameterisation of
\eqn{eq:three_mass_mom_param} gives
\begin{eqnarray}
\left(\alpha_{01}\Bmp{a^-|\s
  K_2|b^-}+t\Bmp{a\Kf1}[\chi
  b]\right).
\end{eqnarray} 
As $S_2=0$ we see that $\alpha_{01}=S_1/\gamma$ and that
$\gamma=2(K_1\cdot K_2)$. Then taking the $t^0$ component of the
$[\Inf_{t}]$ of this in accordance with \eqn{eq:scalar_triangle_MI}
leaves us with our desired coefficient
\begin{eqnarray}
-\frac{S_1}{\Bmp{K_2^-|\s K_1|K_2^-}}\Bmp{a^-|\s
  K_2|b^-},
\end{eqnarray}
which matches the expected result.

\subsubsection{The bubble contributions of a three-mass linear triangle}

Consider a linear triangle with in this case three massive legs, so
now $K_2$ is massive but again $a$ and $b$ are arbitrary massless
four-vectors,
\begin{eqnarray}
-i\int d^4l\frac{\Bmp{a^-|\s l|b^-}}{l^2(l-K_1)^2(l+K_2)^2}.\label{eq:extract_bubble}
\end{eqnarray}
Extracting the bubble coefficient of the integral $B_0(K^2_1)$ is done
by cutting the two propagators $l^2$ and $(l-K_1)^2$. Again cutting
the legs is done by removing the relevant propagators from the
integrand so that it is given by
\begin{eqnarray}
\frac{\Bmp{a^-|\s l|b^-}}{(l+K_2)^2}.\label{eq:example_part_2}
\end{eqnarray}
As this contains a single propagator, and therefore a single pole, we
could choose to set $\chi=\Kf2$ (as defined in
\eqn{eq:def_K2flat_for_bub}), before performing the series expansions
in $y$ and $t$. For this choice of $\chi$ the bubble coefficient comes
entirely from the two-particle cut. Using the first term of
\eqn{eq:scalar_coeff_result} gives directly
\begin{eqnarray}
-i\left(\frac{\gamma\Bmp{a^-|\s {\Kf1}|b^-}}{\gamma^2-S_1S_2}-\frac{S_1\Bmp{a^-|\s {\Kf2}|b^-}}{\gamma^2-S_1S_2}\right),
\label{eq:bble_res_12}
\end{eqnarray}
where $\gamma=\Bmp{\Kfm2|\s {\Kf1}|\Kfm2}$, a result which is
equivalent to the expected answer. 

In order to demonstrate the procedure of using triple cut
contributions in extracting a bubble coefficient we will now reproduce
this by assuming $\chi\neq\Kf2$. For this case the first term of
\eqn{eq:scalar_coeff_result} then gives
\begin{eqnarray}
-i\frac{\Bmp{a\chi}[\Kf1 b]}{\Bmp{\chi^-|\s K_2|\Kfm1}},
\end{eqnarray}
which upon choosing $\chi=a$ vanishes and so the complete contribution
will come from the triple cut pieces of
\eqn{eq:extract_bubble}. Cutting the remaining propagator in
\eqn{eq:example_part_2} gives us the single triple cut term which will
contribute. The integrand of this is given, after multiplying by an
additional factor of $i$ which would come from the third tree
amplitude if this was a triple cut, by
\begin{eqnarray}
&&\hspace*{-0.8cm}
i\left(\Bmp{al}[lb]\Big|_{y=y_{+}}+\Bmp{al}[lb]\Big|_{y=y_{-}}\right)
=
i\left((y_{+}+y_{-})\Bmp{a\Kf1}[\Kf1b]+2t\Bmp{a\Kf1}[a b]\right),\label{eq:3_mass_examp_1}
\end{eqnarray}
where we have set $\chi=a$. From \eqn{eq:def_of_y_for_triple_cut} we
have
\begin{eqnarray}
y_{+}+y_{-}=\frac{\gamma\left(\Bmp{\Kfm1|\s
    K_2|\Kfm1}-S_1\Bmp{a^-|\s K_2|a^-}\right)t+S_1\Bmp{a^-|\s
    K_2|\Kfm1}}{S_1\Bmp{a^-|\s
    K_2|\Kfm1}},
\end{eqnarray}
Hence taking the $[\Inf_{t}]$ of the cut integrand,
\eqn{eq:3_mass_examp_1}, and dropping any terms not proportional to
$t$ leaves
\begin{eqnarray}
it \Bmp{a^-|\s K_1|b^-}\left(\frac{\gamma\Bmp{\Kfm1|\s
    K_2|\Kfm1}-S_1\Bmp{a^-|\s K_2|a^-}}{S_1\Bmp{a^-|\s
    K_2|\Kfm1}}+2\frac{[ab]}{[\Kf1 b]}\right),
\end{eqnarray}
which after inserting the result for the $t$ integral given by
\eqn{eq:the_triangle_t_ints} and substituting this into the second
term of \eqn{eq:scalar_coeff_result} gives for our desired coefficient
\begin{eqnarray}
&&\hspace*{-0.8cm} -i\frac{\Bmp{a^-|\s
  K_1|b^-}}{4\,\Delta}\left(\Bmp{\Kfm1|\s
  K_2|\Kfm1}\!-\!\frac{S_1}{\gamma}\Bmp{a^-|\s
  K_2|a^-}+\frac{2S_1}{\gamma}\Bmp{a^-|\s K_2|\Kfm1}\frac{[a
  b]}{[\Kf1 b]}\right)
\nonumber\\
&=&-i\frac{1}{2\Delta}\left((K_1\cdot K_2)\Bmp{a^-|\s K_1|b^-}-S_1\Bmp{a^-|\s K_2|b^-}\right),
\end{eqnarray}
where $\Delta$ was given in \eqn{eq:def_delta}.  This matches both the
expected result and \eqn{eq:bble_res_12}.

\subsection{Constructing the one-loop six-photon
  amplitude $A_6(1^-,2^+,3^-,4^+,5^-,6^+)$}

Recently an analytic form for the last unknown six-photon one-loop
amplitude was obtained by Binoth, Heinrich, Gehrmann and Mastrolia in
ref.~\cite{Binoth:2007ca}. This result was used to confirm a previous
numerical result~\cite{Nagy:2006xy}. More recently still further
corroboration has been provided by~\cite{Ossola:2007bb}. Here we
reproduce, as an example, the calculation of the three-mass triangle
and bubble coefficients, again confirming part of these results.

\begin{figure}[ht]
\centerline{\epsfxsize 2 truein\epsfbox{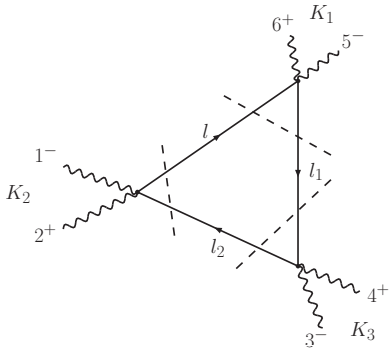}}
\caption{Triple cut six-photon amplitude in the $12:34:56$ channel.
}
\label{figure:triple_cut_photon_12_34_56}
\end{figure}

Firstly it is a very simple exercise to demonstrate by explicit
computation that all bubble coefficients vanish. If we were to use the
basis of finite box integrals, as defined in~\cite{Britto:2005ha},
then there is only a single unique three-mass triangle coefficient, a
complete explicit derivation of which we now present. Starting from
the cut in the $12:34:56$ channel shown in
figure~\ref{figure:triple_cut_photon_12_34_56}
we can write the cut integrand as
\begin{eqnarray}
&&\hspace*{-0.8cm} 16A_4(-l^{-h}_{q},1^-,2^+,l^{h_2}_{2,\qbd})
A_4(-l^{-h_2}_{2,q},3^-,4^+,l^{h_1}_{1,\qbd})A_4(-l^{-h_1}_{1,q},5^-,6^+,l^{h}_{\qbd}),\label{eq:photon_trpl_cut}
\end{eqnarray}
with all unlabelled legs photons and $l_1=l-K_{56}$ and
$l_2=l+K_{12}$. The overall factor of $16$ comes from the differing
normalisation conventions between QCD colour-ordered amplitudes and
QED photon amplitudes. Both helicity choices $h=h_1=h_2=\pm$ give
identical contributions. Written explicitly, \eqn{eq:photon_trpl_cut}
is
\begin{eqnarray}
&&\hspace*{-0.8cm}32i
\frac{\Bmp{l1}^2\Bmp{l_23}^2\Bmp{l_15}^2}{\Bmp{l2}\Bmp{l_22}\Bmp{l_14}\Bmp{l_24}\Bmp{l6}\Bmp{l_16}}.
\end{eqnarray}
After inserting the momentum parameterisation of
\eqn{eq:three_mass_mom_param} this becomes
\begin{eqnarray}
&&\hspace*{-0.8cm}32i
\frac{1}{\left(t\Bmp{\Kf12}+\alpha_{01}\Bmp{\Kf22}\right)\left(t\Bmp{\Kf12}+\alpha_{21}\Bmp{\Kf22}\right)\left(t\Bmp{\Kf14}+\alpha_{11}\Bmp{\Kf24}\right)}
\nonumber\\
&&\hphantom{dt}\times\frac{\left(t\Bmp{\Kf11}+\alpha_{01}\Bmp{\Kf21}\right)^2\left(t\Bmp{\Kf13}+\alpha_{21}\Bmp{\Kf23}\right)^2\left(t\Bmp{\Kf15}+\alpha_{11}\Bmp{\Kf25}\right)^2}{\left(t\Bmp{\Kf14}+\alpha_{21}\Bmp{\Kf24}\right)\left(t\Bmp{\Kf16}+\alpha_{01}\Bmp{\Kf26}\right)\left(t\Bmp{\Kf16}+\alpha_{11}\Bmp{\Kf26}\right)}.
\end{eqnarray}
Applying \eqn{eq:scalar_triangle_MI} implies taking only the $t^0$
piece of the $[\Inf_{t}]$ of this expression. Averaging over both
solutions leaves us with our form for the three mass triangle
coefficient
\begin{eqnarray}
-16i\sum_{\gamma=\gamma_{\pm}}\frac{\Bmp{\Kf11}^2\Bmp{\Kf13}^2\Bmp{\Kf15}^2}{\Bmp{\Kf12}^2\Bmp{\Kf14}^2\Bmp{\Kf16}^2},
\end{eqnarray} 
where $\Kf1$ depends upon the form of $\gamma_{\pm}$ as given
in~\eqn{eq:def_gamma_pm}. Numerical comparison with the analytic
result of~\cite{Binoth:2007ca} shows complete agreement.

\subsection{Contributions to the one-loop
  $A_6(1_{q}^+,2_{\qbd}^-,3^-,4^+;5_{\ebd}^-,6^+_{e})$ amplitude}

This particular amplitude was originally obtained by Bern, Dixon and
Kosower in~\cite{Bern:1997sc}. Making up this amplitude are many box,
triangle and bubble integrals along with rational terms. Here we will
recompute one particular representative three-mass triangle
coefficient in order to highlight the application of our technique to
a phenomenologically interesting process.

\begin{figure}[ht]
\centerline{\epsfxsize 3 truein\epsfbox{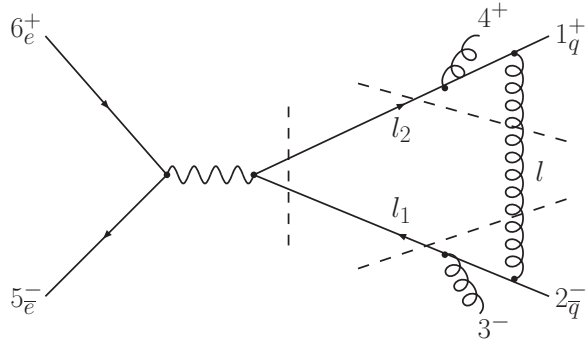}}
\caption{
Triple cut in the $14:23:56$ channel.
}
\label{figure:triple_cut_in14_23_56}
\end{figure}

Following the notation of~\cite{Bern:1997sc}, we wish to calculate the
three-mass triangle coefficient of
$I_{3}^{3m}(s_{14},s_{23},s_{56})\equiv C_0(s_{14},s_{56})$ of the
$F^{cc}$ term. The only contributing cut is shown in figure
\ref{figure:triple_cut_in14_23_56}.
We begin by writing down the triple cut integrand for this case
\begin{eqnarray}
&&\hspace*{-0.8cm}
A_4(-l^{-h_1}_{1,\qb},5_{\ebd}^-,6_e^+,l^{h_2}_{2,q})
A_4(-l^{-h_2}_{2,\qbd},4^+,1_{q}^+,l^{h}_{g})A_4(-l^{-h}_{g},2_{\qbd}^-,3^-,l^{h_1}_{1,q}),
\end{eqnarray}
where $l_1=l-K_{23}$ and $l_2=l+K_{14}$. Only when $h=-$, $h_1=+$ and
$h_2=+$ do we get a contribution. It can be written explicitly as
\begin{eqnarray}
i\frac{\Bmp{l_{2}5}^2
\Bmp{l l_{2}}^2\Bmp{23}^2
}{\Bmp{14}\Bmp{56}\Bmp{4l_{2}}\Bmp{2l}\Bmp{ll_{1}}\Bmp{l_{1}l_{2}}}.
\end{eqnarray}
Rewriting this in terms of the loop momentum parametrisation of
\eqn{eq:three_mass_mom_param} gives
\begin{eqnarray}
i\frac{\gamma\left(t\Bmp{\Kf1 5}+\alpha_{21}\Bmp{\Kf2 5}\right)^2
\Bmp{23}^2
}{s_{23}\left(1-\frac{s_{23}}{\gamma}\right)\Bmp{14}\Bmp{56}\left(t
    \Bmp{4\Kf1}+\alpha_{21}\Bmp{4\Kf2}\right)\left(t \Bmp{2\Kf1}+\alpha_{01}\Bmp{2\Kf2}\right)}\label{eq:3_mass_start_1}.
\end{eqnarray}
The two solutions of $\gamma$ are given by
$\gamma_{\pm}=-(K_{23}\cdot K_{14})\pm\sqrt{(K_{23}\cdot
K_{14})^2-s_{23}s_{14}}$, the $\alpha_{ij}$'s are given in
Appendix~\ref{appendix:the_triangle_param}.

The application of \eqn{eq:scalar_triangle_MI} involves taking
$[\textrm{Inf}_t]$ of \eqn{eq:3_mass_start_1}, dropping all but the
$t^0$ component of the result and then averaging over both solutions
of $\gamma$ giving the coefficient
\begin{eqnarray}
&&\hspace*{-0.8cm}-\frac{i}{2}\sum_{\gamma=\gamma_{\pm}}\frac{\gamma\Bmp{\Kf1 5}^2
\Bmp{23}^2
}{S_1\left(1-\frac{S_1}{\gamma}\right)\Bmp{14}\Bmp{56}\Bmp{4\Kf1}\Bmp{2\Kf1}},\label{eq:3_mass_inf_1}
\end{eqnarray}
where again $\Kf1$ depends upon $\gamma_{\pm}$.  Numerical comparison
against the solution for this coefficient presented
in~\cite{Bern:1997sc},
\begin{eqnarray}
-\frac{i}{2}\frac{[14]\left(\Bmp{2^-|\s K_{14}\s
 K_{23}|5^+}^2-\Bmp{25}^2s_{14}s_{23}\right)}{\Bmp{14}[23]\Bmp{56}\Bmp{2^-|\s
 K_{14}|3^-}\Bmp{2^-|\s K_{34}|1^-}}+\textrm{flip},
\end{eqnarray} 
shows complete agreement, where the operation $\textrm{flip}$ is
defined as the exchanges $1\leftrightarrow 2$, $3\leftrightarrow 4$,
$5\leftrightarrow 6$, $\Bmp{ab}\leftrightarrow [ab]$.

The remaining triangle and bubble coefficients can be derived in an
analogous way. We have computed a selection of these coefficients for
$A_6(1_{q}^+,2_{\qbd}^-,3^-,4^+;5_{\ebd}^-,6^+_{e})$, along with
coefficients of other amplitudes given in~\cite{Bern:1997sc}, and find
complete agreement.

\subsection{Bubble coefficients of the one-loop 5-gluon QCD amplitude $A_5(1^-,2^-,3^+,4^+,5^+)$}

This result for the 1-loop 5 gluon QCD amplitude
$A_5(1^-,2^-,3^+,4^+,5^+)$ was originally calculated by Bern, Dixon,
Dunbar and Kosower in~\cite{Neq4Oneloop2}. It contains neither box nor
triangle integrals, only bubbles. We need therefore only compute
bubble coefficients. There are only a pair of such coefficients, with
masses $s_{23}$ and $s_{234}=s_{51}$.

For the first cut in the channel $K_1=K_{23}$ we have, for the sum of
the two possible helicity configurations, the two-particle cut
integrand
\begin{eqnarray}
\frac{2}{\Bmp{23}\Bmp{45}\Bmp{51}}\frac{\Bmp{1l_1}^2\Bmp{1l}\Bmp{2l}\Bmp{2l_1}^2}{\Bmp{4l_1}\Bmp{3l_1}\Bmp{ll_1}^2},
\end{eqnarray}
and for the second, in the channel $K_1=K_{234}$,
\begin{eqnarray}
\frac{2}{\Bmp{23}\Bmp{34}\Bmp{51}}\frac{\Bmp{1l_1}^2\Bmp{1l}\Bmp{2l}\Bmp{2l_1}^2}{\Bmp{4l_1}\Bmp{5l_1}\Bmp{ll_1}^2}.
\end{eqnarray}

Focus upon the $K_1=K_{23}$ cut initially. There are two
pole-containing terms in the denominator of this cut. We could choose
to partial fraction these terms and then pick $\chi=K_2$ in each case
to extract the coefficient. Instead though we will derive the
coefficient using triple cut contributions. Choosing $\chi=k_1$ so
that after inserting the cut loop momentum parameterisation of
\eqn{eq:dble_cut_mom_def} the cut integrand becomes
\begin{eqnarray}
&&\hspace*{-0.8cm}\frac{2\gamma^2\Bmp{1\Kf1}}{S_1^2\Bmp{23}\Bmp{45}\Bmp{51}}
t
\frac{\left(\Bmp{2\Kf1}-\frac{S_1}{\gamma}\frac{y}{t}\Bmp{21}\right)\left(t\Bmp{2\Kf1}+\frac{S_1}{\gamma}\left(1-y\right)\Bmp{21}\right)}{\left(\Bmp{3\Kf1}-\frac{S_1}{\gamma}
  \frac{y}{t}\Bmp{31}\right)\left(\Bmp{4\Kf1}-\frac{S_1}{\gamma}
  \frac{y}{t}\Bmp{41}\right)},
\end{eqnarray}
and hence produces no $[\Inf_{y} [\Inf_{t}]]$ term. Consequentially
the two-particle cut contribution to the bubble coefficient
vanishes. The same choice of $\chi$ similarly removes all two-particle
cut contributions in the channel $K_1=K_{234}$ from the corresponding
scalar bubble coefficient.

Examining the triple cuts of the bubble in the $K_{23}$ channel shows
only two possible contributions, again after summing over both
contributing helicities, given by
\begin{eqnarray}
\frac{2i}{\Bmp{45}\Bmp{51}}\frac{[3l][3l_2]\Bmp{1l_1}\Bmp{1l}^2\Bmp{2l_1}\Bmp{2l}}{\Bmp{ll_1}\Bmp{l_1l_2}[ll_2]\Bmp{l4}},
\end{eqnarray}
when $K_2=k_3$ and
\begin{eqnarray}
-\frac{2i}{\Bmp{23}\Bmp{51}}\frac{[4l][4l_2]\Bmp{1l_1}\Bmp{1l_2}^2\Bmp{2l_1}\Bmp{2l}^2}{\Bmp{ll_1}\Bmp{l_1l_2}[ll_2]\Bmp{5l_2}\Bmp{3l}},
\end{eqnarray}
when $K_2=k_4$. In both cases $K_2$ is massless and is of positive
helicity so we use the parameterisation of the triple cut momenta for
$y_+$ given in \eqn{eq:y_+_two_mass_sol}. Then along with setting
$\chi=k_1$ gives for the first triple cut integrand
\begin{eqnarray}
\frac{2i\Bmp{1\Kf1}^2\Bmp{23}}{\Bmp{13}\Bmp{34}\Bmp{45}\Bmp{51}}\frac{\gamma^2}{S_1^2}t\left(\frac{\Bmp{1^-|\s 2|3^-}}{\Bmp{1\Kf1}}\!-\!\frac{\gamma
t}{S_1}\frac{\Bmp{3^-|\s
K_{23}|3^-}}{\Bmp{13}}\right)\!\!\left(t\frac{\Bmp{1\Kf1}}{\Bmp{13}}\Bmp{23}\!+\!\frac{S_1}{\gamma}\Bmp{21}\right),
\end{eqnarray}
and for the second
\begin{eqnarray}
-\frac{2i\Bmp{1\Kf1}^2\Bmp{24}^2}{\Bmp{23}\Bmp{34}\Bmp{45}\Bmp{51}\Bmp{14}}\frac{\gamma^2}{S_1^2}t\!\left(\!\frac{\gamma t}{S_1}\frac{\Bmp{4^-|\s
      K_{23}|4^-}}{\Bmp{1
      4}}\!-\!\frac{\Bmp{1^-|\s K_{23}|4^-}}{\Bmp{1\Kf1}}\!\right)\!\!\left(\!t\frac{\Bmp{1\Kf1}}{\Bmp{14}}\Bmp{24}\!+\!\frac{S_1\Bmp{21}}{\gamma}\!\right).
\end{eqnarray}
Applying these integrands to the second term of
\eqn{eq:scalar_coeff_result} by taking $[\Inf_{t}]$, dropping any
terms not proportional to $t$ and then performing the substitution
$t^i\rightarrow T(i)$ gives for the coefficient of the first triple
cut simply $\frac{1}{3}\At_5$, and for the second triple cut
\begin{eqnarray}
-\frac{\At_5}{s_{12}^3}\frac{\Bmp{1^+|\s 2\s 4\s K_{23}|1^+}^2}{\Bmp{4^-|\s
	K_{23}|4^-}^2}\left(
s_{12}
-\frac{2}{3}\frac{\Bmp{1^+|\s 2 \s 4\s K_{23}|1^+}}{\Bmp{4^-|\s
	K_{23}|4^-}}\right).
\end{eqnarray}
After following the same series of steps as above for the second
bubble coefficient with $K_1=K_{234}$ we find only a single triple cut
contributing term corresponding to $K_2=k_4$. This is related to the
second triple cut coefficient derived above via the replacement
$K_{23}\rightarrow K_{234}$ and swapping the overall sign.

After combining the three triple cut pieces above we arrive at the
following form for the cut constructable pieces of this amplitude
\begin{eqnarray}
&&\hspace*{-0.8cm}
\frac{r_{\Gamma}(\mu^2)^{\epsilon}}{(4\pi)^{2-\epsilon}}\Bigg(\frac{1}{3}\At_5 B_0(s_{23})
\nonumber\\
&&\hphantom{12345}-\frac{\At_5}{s_{12}^3}
\frac{\Bmp{1^+|\s 2\s 4\s K_{23}|1^+}^2}{\Bmp{4^-|\s
	K_{23}|4^-}^2}\!\left(
\!s_{12}\!
-\!\frac{2}{3}\frac{\Bmp{1^+|\s 2 \s 4\s K_{23}|1^+}}{\Bmp{4^-|\s
	K_{23}|4^-}}\!\right)\!
\left(B_0(s_{23})\!-\!B_0(s_{234})\right)\Bigg),
\end{eqnarray}
which can easily be shown to match the result given
in~\cite{Neq4Oneloop2}.

While this example is particularly simple we have also performed
additional comparisons against other results in the literature. Such
tests include the cut constructible pieces of all two-minus gluon
amplitudes with up to seven external legs, originally obtained in
~\cite{Neq4Oneloop2,BBSTQCD}. Additionally we find agreement for the
case when, with six gluon legs, three are of negative helicity and
adjacent to each other and the remainder are positive helicity, which
was originally obtained in~\cite{Bern:2005hh}. We have also
successfully reproduced the known three mass triangle coefficients in
$\mathcal{N}=1$ supersymmetry for $A_6(1^-,2^+,3^-,4^+,5^-,6^+)$ and
$A_6(1^-,2^-,3^+,4^-,5^+,6^+)$, originally obtained
in~\cite{Britto:2005ha}.



\section{Conclusions}
\label{ConclusionSection}

The calculation of Standard Model background processes at the LHC
requires efficient techniques for the production of amplitudes. The
large numbers of processes involved along with their differing
partonic makeups suggests that as much automation as possible is
desired. In this paper we have presented a new formalism which directs
us towards this goal. Coefficients of the basis scalar integrals
making up a one-loop amplitude are constructed in a straightforward
manner involving only a simple change of variables and a series
expansion, thus avoiding the need to perform any integration or
calculate any extraneous intermediate quantities. The main results of
this paper can be encapsulated simply by \eqn{eq:scalar_triangle_MI}
and \eqn{eq:scalar_coeff_result} along with the cut loop momentum
given by \eqn{eq:three_mass_mom_param}, \eqn{eq:dble_cut_mom_def} and
\eqn{eq:def_of_y_for_triple_cut}.

Although this technique has been presented mainly in the context of
using generalised
unitarity~\cite{Bern:1997sc,TwoLoopSplit,NeqFourSevenPoint,Eden} to
construct coefficients, and hence the cut-constructible part of the
amplitude, it can also be used as an efficient method of performing
one-loop integration. Using the idea of ``cutting'' two, three or four
of the propagators inside an integral, we isolate and then extract
scalar basis coefficients. This procedure then allows us to rewrite
the integral in terms of the scalar one-loop basis integrals, hence
giving us a result for the integral.

Different unitarity cuts isolate particular basis integrals. For the
extraction of triangle integral coefficients this means triple cuts
and for bubble coefficients we use a combination of two-particle and
triple cuts. Extracting the desired coefficients from these cut
integrands is then a two step process. The first step is to rewrite
the cut loop momentum in terms of a parameterisation which depends
upon the remaining free parameters of the integral after all the cut
delta functions have been applied. Triangle coefficients are then
found by taking the terms independent of the sole free integral
parameter as this parameter is taken to infinity.

Bubble coefficients are calculated in a similar if slightly more
complicated way. The presence of a second free parameter in the bubble
case means that we must take into account, not only the constant term
in the expansion of the cut integrand as the free integral parameters
are taken to infinity, but also powers of one of these parameters. The
limit on the maximum power of $l^{\mu}$ appearing in the cut integral
restricts the appearance of such terms and hence we need consider only
finite numbers of powers of these free parameters. Additionally it can
also be necessary to take into account contributions from terms
generated by applying an additional cut to the bubble integral. The
flexibility in our choice of the cut-loop momentum parameterisation
allows us to directly control whether we need compute any of these
triple cut terms. Furthermore we can control which of these triple cut
terms appears, in cases when their computation is necessary.

As we consider the application of this procedure to more diverse
processes than those detailed here, we should also investigate the
``complexity'' of the generated coefficients. In the applications we
have presented we can see that we produce ``compact'' forms with
minimal amounts of simplification required.  This is important if we
are to consider further automation. The straightforward nature of this
technique combined with the minimal need for simplification means that
efficient computer implementations can easily be produced.  As a test
of this assertion we have implemented the formalism within a {\tt
Mathematica} program which has been used to perform checks against
state-of-the-art results contained in the literature. Such checks have
included various helicity configurations of up to seven external
gluons as well as the bubble and three-mass triangle coefficients of
the six photon $A_6(-+-+-+)$ amplitude. In addition representative
coefficients of processes of the type $e^+e^-\rightarrow
q\overline{q}gg$ have been successfully obtained.

Our procedure as presented has mainly been in the context of massless
theories. Fundamentally there is no restriction to the application of
this to theories also involving massive fields circulating in the
loop. Extensions to include masses should require only a suitable
momentum parameterisation for the cut loop momentum; the procedure is
then expected to apply as before.

In conclusion therefore we believe that the technique presented here
shows great potential for easing the calculation of needed one-loop
integrals for current and future colliders.

\section*{Acknowledgements}

I would like to thank David Kosower for collaboration in the early
stages of this work and also Zvi Bern and Lance Dixon for many
interesting and productive discussions as well as for useful comments
on this manuscript. I would also like to thank the hospitality of
Saclay where early portions of this work were carried out. The figures
were generated using Jaxodraw~\cite{Jaxo}, based on
Axodraw~\cite{Axo}.


\appendix


\section{The triple cut parameterisation}
\label{appendix:the_triangle_param}

In this appendix we give the complete detail of the triple cut
parameterisation along with some other useful results.
The three cut momenta are given by
\begin{eqnarray}
\langle l_i^-|
=t\langle K_{1}^{\flat,-}|+\alpha_{i1}\langle K_{2}^{\flat,-}|,
&&
\langle l_i^+|
=\frac{\alpha_{i2}}{t}\langle K_{1}^{\flat,+}|+\langle K_{2}^{\flat,+}|,
\end{eqnarray}
with
\begin{eqnarray}
\alpha_{01}=\frac{S_1\left(\gamma-S_2\right)}{\left(\gamma^2-S_1S_2\right)},\;\;&&\;\;\alpha_{02}=\frac{S_2\left(\gamma-S_1\right)}{\left(\gamma^2-S_1S_2\right)},
\nonumber\\
\alpha_{11}=\alpha_{01}-\frac{S_1}{\gamma}=-\frac{S_1S_2\left(1-(S_1/\gamma)\right)}{\gamma^2-S_1S_2},\;\;&&\;\;
\alpha_{12}=\alpha_{02}-1=\frac{\gamma(S_2-\gamma)}{\gamma^2-S_1S_2},
\nonumber\\
\alpha_{21}=\alpha_{01}-1=\frac{\gamma(S_1-\gamma)}{\gamma^2-S_1S_2},\;\;&&\;\;
\alpha_{22}=\alpha_{02}-\frac{S_2}{\gamma}=-\frac{S_1S_2\left(1-(S_2/\gamma)\right)}{\gamma^2-S_1S_2},
\end{eqnarray}
along with the identities
$\alpha_{01}\alpha_{02}=\alpha_{11}\alpha_{12}$ and
$\alpha_{01}\alpha_{02}=\alpha_{21}\alpha_{22}$.  When written as
four-vectors the cut momentum are given by
\begin{eqnarray}
l^{\mu}_i=\alpha_{i2} K_1^{\flat,\mu}+\alpha_{i1} K_2^{\flat,\mu}+\frac{t}{2}\Bmp{K^{\flat,-}_1|\gamma^{\mu}|K^{\flat,-}_2}
+\frac{\alpha_{i1}\alpha_{i2}}{2t}\Bmp{K^{\flat,-}_2|\gamma^{\mu}|K^{\flat,-}_1}.\label{eq:sol_for_l_in_3_mass_i}
\end{eqnarray}

From these parameterised forms we have the following spinor product
identities
\begin{eqnarray}
\left[ll_1\right]&=&\frac{\alpha_{12}-\alpha_{02}}{t}[K^{\flat}_1
  K^{\flat}_2]=-\frac{1}{t}[K^{\flat}_2
  K^{\flat}_1],
\nonumber\\
\Bmp{ll_1}&=&t(\alpha_{11}-\alpha_{01})\Bmp{K^{\flat}_1
  K^{\flat}_2}=-\frac{tS_1}{\gamma}\Bmp{K^{\flat}_1 K^{\flat}_2},
\nonumber\\
\left[ll_2\right]&=&\frac{\alpha_{22}-\alpha_{02}}{t}[K^{\flat}_1
  K^{\flat}_2]=-\frac{S_2}{\gamma t}[K^{\flat}_2
  K^{\flat}_1],
\nonumber\\
\Bmp{ll_2}&=&t(\alpha_{21}-\alpha_{01})\Bmp{K^{\flat}_1
  K^{\flat}_2}=-t\Bmp{K^{\flat}_1 K^{\flat}_2},
\nonumber\\
\left[l_1l_2\right]&=&\frac{\alpha_{22}-\alpha_{12}}{t}[K^{\flat}_1K^{\flat}_2]=\frac{1}{t}\left(1-\frac{S_2}{\gamma}\right)[K^{\flat}_2K^{\flat}_1],
\nonumber\\
\Bmp{l_1l_2}&=&t(\alpha_{11}-\alpha_{21})\Bmp{K^{\flat}_1K^{\flat}_2}=-t\left(1-\frac{S_1}{\gamma}\right)\Bmp{K^{\flat}_1K^{\flat}_2}.
\label{eq:the_spinor products_res}
\end{eqnarray}
and we note that 
\begin{eqnarray}
-\left(1-\frac{S_2}{\gamma}\right)\left(1-\frac{S_1}{\gamma}\right)\gamma=-\gamma-\frac{S_1S_2}{\gamma}+S_1+S_2=(K_1-K_2)^2=S_3,
\end{eqnarray}
and so with $l\equiv l_0$ we have $\Bmp{l_il_j}[l_jl_i]=S_{i+j}$, as
expected.


\section{The triple cut bubble contribution momentum parameterisation when $K_2^2=0$} 
\label{appendix:forced_triple_cut_S_2_0}

In this appendix we give the forms for the triple cut momentum of
section~\ref{section:forced_triple_mom} in the case when $S_2=0$,
i.e. we have a one or two mass triangle. Firstly in these cases the
$K_2$ leg is attached to a three-point vertex and so the amplitude for
this will contain either $[K_2l]$ or $\Bmp{K_2l}$ depending upon the
helicity of $K_2$. This means that in the positive helicity case only
the delta function solution $\delta(y-y_{+})$ survives and for a
negative helicity $K_2$ the $\delta(y-y_{-})$ survives. We have for
both solutions
\begin{eqnarray}
&&\hspace*{-0.8cm}J'_t=
  \frac{1}{\left(S_1\Bmp{\chi^-|\s K_2|\Kfm1}+t \gamma \Bmp{K_2^-|\s K_1|K_2^-}\right)}.
\label{eq:integrand_for_a_scalar triangle}
\end{eqnarray}

The momentum parameterisation for the $y_{+}$ solution is given in
spinor components by
\begin{eqnarray}
\langle l^-|
=\frac{\Bmp{\chi \Kf1}}{\Bmp{\chi K_2}}\langle
K_{2}^{-}|,\;
&&
\;\langle l^+|=\langle
K_{1}^{\flat,+}|-\frac{\gamma t}{S_1}\frac{1}{\Bmp{\chi K_2}}\langle
K_2^-|\s K_1,\label{eq:y_+_two_mass_sol}
\end{eqnarray}
and as a 4-vector by
\begin{eqnarray}
l^{\mu}=\frac{\Bmp{\chi \Kf1}}{2\Bmp{\chi K_2}}\left(\frac{t\gamma}{S_1\Bmp{\chi K_2}}\Bmp{K_2^-|\gamma^{\mu}\s K_1 |K_2^+}+\Bmp{K_2^-|\gamma^{\mu}|\Kfm1}\right).\label{eq:forced_triple_yminus_sol_l_+}
\end{eqnarray}
The other momenta are given by
\begin{eqnarray}
\langle l_1^-|=t\frac{\Bmp{\chi \Kf1}}{\Bmp{\chi K_2}}\langle
K_{2}^{-}|-\frac{S_1}{\gamma}\langle
\chi^{-}|,\;\;&&\;\;
\langle l_1^+|=-\frac{\gamma}{S_1\Bmp{\chi K_2}}\langle K_2^-|\s K_1,
\nonumber\\
\langle l_2^-|=\frac{\Bmp{\chi \Kf1}}{\Bmp{\chi K_2}}\langle
K_{2}^{-}|,\;\;&&\;\;
\langle l_2^+|=-\frac{\langle \chi^-|\s K_3}{\Bmp{\chi \Kf1}}-\frac{\gamma t}{S_1}\frac{1}{\Bmp{\chi K_2}}\langle
K_2^-|\s K_1.
\end{eqnarray}
where we have moved the overall factor of $t$ from $\langle l_1^-|$ to
$\langle l_1^+|$ to avoid the presence of a $1/t$ term for aesthetical
reasons.  The spinor products formed from these are given by
\begin{eqnarray}
\Bmp{ll_1}=\frac{S_1}{\gamma}\Bmp{\chi
  \Kf1},\;\;&&\;\;[ll_1]=[\Kf1 \chi],
\nonumber\\
\Bmp{ll_2}=0,\;\;&&\;\;[ll_2]=-\frac{\Bmp{\chi K_2}}{\Bmp{\chi \Kf1}}[lK_2],
\nonumber\\
\Bmp{l_1l_2}=-\frac{S_1}{\gamma}\Bmp{\chi
  \Kf1},\;\;&&\;\;[l_1l_2]=-\frac{S_3}{S_1}[\Kf1\chi].
\end{eqnarray}
and we see that again, as expected, with $l=l_0$, we have
$\Bmp{l_il_j}[l_jl_i]=S_{i+j}$. As we have massless legs some spinor
products will consequentially vanish. In the two-mass case these are
\begin{eqnarray}
\Bmp{ll_2}=0,\;\;\;\;\;\Bmp{l
  K_2}=0,\;\;\;\;\;[l_2K_2]=0,
\end{eqnarray}
and for the one-mass case
\begin{eqnarray}
[l_1l_2]=0,\;\;\;\;\;\Bmp{ll_2}=0,\;\;\;\;\;\Bmp{l
  K_2}=0,\;\;\;\;\;\Bmp{l_2K_2}=0,\;\;\;\;\;[l_1K_3]=0,\;\;\;\;\;[l_2K_3]=0,
\end{eqnarray}
where $K_3$ is the momentum of the third leg.

The momentum parameterisation for the $y_{-}$ solution is given in
spinor components by
\begin{eqnarray}
\langle l^-|
=\frac{t}{[K_2\Kf1]}\langle
K_{2}^{+}|\s K_1+\frac{S_1}{\gamma}\langle
\chi^{-}|,\;\;&&\;\;
\langle l^+|=\frac{[\chi \Kf1]}{[K_2\Kf1]}\langle K_{2}^{+}|,
\end{eqnarray}
and as a 4-vector by
\begin{eqnarray}
l^{\mu}=\frac{[\chi \Kf1]}{2[K_2\Kf1]}\left(\frac{t}{[K_2\Kf1]}\Bmp{K_2^+|\s K_1 \gamma^{\mu}|K_2^-}+\frac{S_1}{\gamma}\Bmp{\chi^-|\gamma^{\mu}|K_2^-}\right).\label{eq:forced_triple_yminus_sol_l}
\end{eqnarray}
The other momenta are given by
\begin{eqnarray}
\langle l_1^-|=\frac{1}{[K_2\Kf1]}\langle
K_2^{+}|\s K_1,\;\;&&\;\;
\langle l_1^+|=-t\frac{[\Kf1 \chi]}{[K_2\Kf1]}\langle K_2^+|-\langle \Kfp1|,
\nonumber\\
\langle
l_2^-|=\frac{1}{[\chi\Kf1]}\langle \Kfp1|\s K_3+\frac{t}{[K_2\Kf1]}\langle K_2^+|\s K_1,\;\;&&\;\;
\langle l_2^+|=\frac{[\chi \Kf1]}{[K_2\Kf1]}\langle K_{2}^{+}|.
\end{eqnarray}
The spinor products formed from these are given by
\begin{eqnarray}
\Bmp{ll_1}=\frac{S_1}{\gamma}\Bmp{\chi
  \Kf1},\;\;&&\;\;[ll_1]=[\Kf1 \chi],
\nonumber\\
\Bmp{ll_2}=\frac{[K_2\Kf1]}{[\chi \Kf1]}\Bmp{K_2l},\;\;&&\;\;[ll_2]=0,
\nonumber\\
\Bmp{l_1l_2}=\frac{S_3}{\gamma}\Bmp{\Kf1\chi},\;\;&&\;\;[l_1l_2]=[\chi\Kf1].
\end{eqnarray}
and again $\Bmp{l_il_j}[l_jl_i]=S_{i+j}$ as expected. The vanishing
spinor products in the two mass case are
\begin{eqnarray}
[ll_2]=0,\;\;\;\;\;\;[lK_2]=0,\;\;\;\;\;\;[l_2K_2]=0,
\end{eqnarray}
and in the one mass case
\begin{eqnarray}
\Bmp{l_1l_2}=0,\;\;\;\;\;[ll_2]=0,\;\;\;\;\;[lK_2]=0,\;\;\;\;\;[l_2K_2]=0,\;\;\;\;\;\Bmp{l_1K_3}=0,\;\;\;\;\;\Bmp{l_2K_3}=0.
\end{eqnarray}

\section{The scalar integral functions}
\label{appendix:scalar_int_fns}

\begin{figure}[ht]
\centerline{\epsfxsize 2 truein\epsfbox{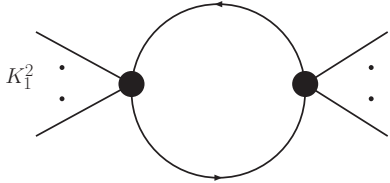}}
\caption{ The scalar bubble integral with a leg of mass $K^2_1$.  }
\label{figure:bubble_def}
\end{figure}

The scalar bubble integral with massive leg $K_1$ given in
figure~\ref{figure:bubble_def}
is defined as
\begin{eqnarray}
B_0(K^2_{1})=(-i)(4\pi)^{2-\epsilon}\int\frac{d^{4-2\epsilon}l}{(2\pi)^{4-2\epsilon}}\frac{1}{l^2(l-K_{1})^2},\label{eq:bubble_int}
\end{eqnarray}
and is given by
\begin{eqnarray}
B_0(K^2_{1})=\frac{r_{\Gamma}}{\epsilon(1-2\epsilon)}(-K_{1}^2)^{-\epsilon}=r_{\Gamma}\left(\frac{1}{\epsilon}-\ln
  (-K_{1}^2)+2\right)+\mathcal{O}(\epsilon),
\end{eqnarray}
with
\begin{eqnarray}
r_{\Gamma}=\frac{\Gamma(1+\epsilon)\Gamma^2(1-\epsilon)}{\Gamma(1-2\epsilon)}.
\end{eqnarray}

The general form of the scalar triangle integral with the masses of
its legs labelled $K_1^2$, $K_2^2$ and $K_3^2$
\begin{figure}[ht]
\centerline{\epsfxsize 2 truein\epsfbox{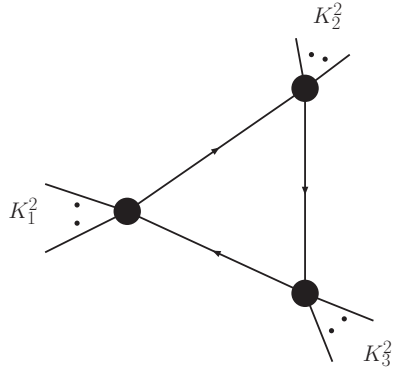}}
\caption{ The scalar triangle with its three legs of mass $K_1^2$,
  $K_2^2$ and $K_3^2$.  }
\label{figure:triangle_diagram}
\end{figure}
given in figure~\ref{figure:triangle_diagram} is defined as 
\begin{eqnarray}
C_0(K^2_1,K^2_2)=i(4\pi)^{2-\epsilon}\int\frac{d^{4-2\epsilon}l}{(2\pi)^{4-2\epsilon}}\frac{1}{l^2(1-K_1)^2(l-K_2)^2},\label{eq:tri_int_def}
\end{eqnarray}
and separates into three cases depending upon the masses of these
external legs. In the one mass case we have $K^2_2=0$ and $K^2_3=0$
and the corresponding integral is given by
\begin{eqnarray}
C_0(K^2_1,K^2_2)=\frac{r_{\Gamma}}{\epsilon^2}(-K^2_1)^{-1-\epsilon}=\frac{r_{\Gamma}}{(-K_1^2)}\left(\frac{1}{\epsilon^2}-\frac{\ln
  (-K_{1}^2)}{\epsilon}+\frac{\ln^2(-K_{1}^2)}{2}\right)+\mathcal{O}(\epsilon),
\end{eqnarray}
If two legs are massive the integral, assuming $K_3^2=0$, is given by
\begin{eqnarray}
C_0(K^2_1,K^2_2)&\!\!=&\!\!\frac{r_{\Gamma}}{\epsilon^2}\frac{(-K^2_1)^{-\epsilon}-(-K^2_2)^{-\epsilon}}{(-K^2_1)-(-K^2_2)}
\nonumber\\
&\!\!=&\!\!\frac{r_{\Gamma}}{(-K_1^2)-(-K_2^2)}\!\left(\!-\frac{\ln\left(-K_1^2\right)-\ln\left(-K_2^2\right)}{\epsilon}\!+\!\frac{\ln^2\left(-K_1^2\right)-\ln^2\left(-K_2^2\right)}{2}\!\right).
\end{eqnarray}
Finally if all three legs are massive then the integral is as given
in~\cite{Bern:1993kr,HuPerezSLAC}
\begin{eqnarray}
C_0(K^2_1,K^2_2)=\frac{i}{\sqrt{\Delta_3}}\sum_{j=1}^3\left[\textrm{Li}_2\left(-\left(\frac{1+i
  \delta_j}{1-i\delta_j}\right)\right)-\textrm{Li}_2\left(-\left(\frac{1-i
  \delta_j}{1+i\delta_j}\right)\right)\right]+\mathcal{O}(\epsilon),
\end{eqnarray}
where
\begin{eqnarray}
\delta_1=\frac{K_1^2-K_2^2-(K_1+K_2)^2}{\sqrt{\Delta_3}},\nonumber\\
\delta_1=\frac{-K_1^2+K_2^2-(K_1+K_2)^2}{\sqrt{\Delta_3}},\nonumber\\
\delta_1=\frac{-K_1^2-K_2^2+(K_1+K_2)^2}{\sqrt{\Delta_3}},
\end{eqnarray}
and
\begin{eqnarray}
\Delta_3=-(K_2^2)^2-(K_2^2)^2-(K_3^2)^2+2(K_1^2K_2^2+K_3^2K_1^2+K_2^2K_3^2)=-4\Delta,
\end{eqnarray}
with $\Delta$ given by \eqn{eq:def_delta}.

The general form for a scalar box function is given by
\begin{eqnarray}
D_0(K^2_1,K^2_2,K^2_3)=(-i)(4\pi)^{2-\epsilon}\int\frac{d^{4-2\epsilon}l}{(2\pi)^{4-2\epsilon}}\frac{1}{l^2(l-K_1)^2(l-K_2)^2(l-K_3)^2}.
\end{eqnarray}
The solution of this integral is split up into classes depending upon
the masses of the external legs. These solutions are labelled as zero
mass $I^{0m}_4$, one mass $I^{1m}_4$, two mass hard, $I_4^{2mh}$, two
mass easy $I_4^{2me}$, three mass $I_4^{3m}$ and four mass $I_4^{4m}$
integrals. The results for which can be found in the literature, for
example in~\cite{Neq4Oneloop}.


\end{document}